\documentclass[aip,notitlepage, floatfix, jcp, preprint]{revtex4-1}%
\usepackage{amssymb}
\usepackage{amsfonts}
\usepackage{amsmath}
\usepackage{graphicx}
\usepackage{enumerate}
\usepackage[usenames,dvipsnames]{color}
\usepackage{ulem}%
\providecommand{\U}[1]{\protect\rule{.1in}{.1in}}

\begin{document}
\title{Classical nucleation theory from a dynamical approach to nucleation}
\author{James F. Lutsko}
\affiliation{Center for Nonlinear Phenomena and Complex Systems, Code Postal 231,
Universit\'{e} Libre de Bruxelles, Blvd. du Triomphe, 1050 Brussels, Belgium}
\email{jlutsko@ulb.ac.be}
\homepage{http://www.lutsko.com}
\author{Miguel A. Dur\'{a}n-Olivencia}
\affiliation{\label{IACT}Laboratorio de Estudios Cristalogr\'{a}ficos. Instituto Andaluz de
Ciencias de la Tierra. CSIC-UGR, Avda. de las Palmeras, 18100 Granada, Spain}
\email{maduran@lec.csic.es}

\begin{abstract}
It is shown that diffusion-limited classical nucleation theory (CNT) can be
recovered as a simple limit of the recently proposed dynamical theory of
nucleation based on fluctuating hydrodynamics (Lutsko, JCP 136, 034509
(2012)). The same framework is also used to construct a more realistic theory
in which clusters have finite interfacial width. When applied to the dilute
solution/dense solution transition in globular proteins, it is found that the
extension gives corrections to the the nucleation rate even for the case of
small supersaturations due to changes in the monomer distribution function and
to the excess free energy. It is also found that the monomer
attachement/detachment picture breaks down at high supersaturations
corresponding to clusters smaller than about 100 molecules. The results also
confirm the usual assumption that most important corrections to CNT can be
acheived by means of improved estimates of the free energy barrier. The theory
also illustrates two topics that have received considerable attention in the
recent literature on nucleation: the importance sub-dominant corrections to
the capillary model for the free energy and of the correct choice of the
reaction coordinate.

\end{abstract}
\date{\today }
\maketitle

\section{Introduction}

The starting point for any discussion of nucleation is the set of ideas
collectively known as Classical Nucleation Theory (CNT)(see, e.g. Ref.
\onlinecite{Kashchiev}). The aim of CNT is to describe the growth and
dissipation of clusters of a new phase forming within a bath of the mother
phase. (This paper will primarily address the problem of homogeneous
nucleation although the ideas discussed here, and indeed those of CNT, are
easily extended to the case of heterogeneous nucleation.) In CNT, the central
quantity is the concentration, $c_{N}$, of clusters of the new phase of a
given size, $N$. (Note that since we will limit the present discussion to
phase transitions involving a single species, the number of molecules in a
cluster and the mass of the cluster are equivalent and the terms will be used
interchangeably along with the generic term "size".) The basic idea underlying
CNT is to write equations for the time evolution of the cluster concentrations
which are analogous in form to chemical rate equations (the Becker-D\"{o}ring
equation\cite{BeckerDoring, Kashchiev}). The physics of the problem enters via
the rate coefficients which depend on the properties of the cluster (e.g. its
size) as well as on those of the mother phase (e.g. its pressure). These
equations are then used to extract a nucleation rate under the particular
assumption of a quasi-stationary distribution in which there is a constant
concentration of monomers and a steady production of clusters up to the of the
critical cluster: critical clusters are removed and new monomers added thus
allowing for the establishment of a steady state. This process is necessary
since the system is, by definition, not in an equilibrium state and so the
concept of an equilibrium distribution is not available.

One important assumption that is made in CNT, even at this level, is that the
concept of a "cluster" is well defined. Of course, one can always devise a
definition of what constitutes a cluster. For example, in their simulation
studies of nucleation, ten Wolde and Frenkel have defined a liquid cluster
forming in a vapor to be the collection of molecules for which the number of
molecules within a given difference exceeds some threshold - i.e., that the
local density in the neighborhood of a molecule is above some
value\cite{frenkel_gas_liquid_nucleation}. The need for such a definition
highlights the underlying fact that clusters are not physically distinct
objects. Indeed, both simulation\cite{frenkel_gas_liquid_nucleation} and
Density Functional Theory calculations\cite{OxtobyEvans, LutskoBubble2, Ghosh,
EvansArcherNucleation} show, e.g., that a liquid cluster forming in a vapor
has a broad and diffuse interface with the surrounding vapor. In fact, for
small clusters, \emph{all} molecules in the cluster could be classified as
being part of the interface with no bulk material present at all. It is only
when clusters are large, specifically when the radius of the cluster is large
compared to the width of the interface, that the distinction between molecules
being "within" a cluster and "outside" a cluster makes sense. However, this is
not a particular problem in the context of CNT as the assumption of large
clusters is usually made at some point in any case.

Another subtlety arises in the description of small clusters. Let us consider
for a moment an \emph{under-saturated} solution for which there is no
difficulty in assuming an equilibrium distribution of cluster sizes. In CNT it
is usually assumed that the probability density for observing a cluster of
size $N$ is proportional to the Boltzmann factor $P\left(  N\right)  \sim
\exp\left(  -\beta\Omega\left(  N\right)  \right)  $ where $\beta=1/k_{B}T$,
$k_{B}$ is Boltzmann's constant, $T$ is the temperature and $\Omega\left(
N\right)  $ is the free energy (e.g. in the capillary approximation) of a
cluster of size (mass) $N$. On the other hand, in the capillary approximation,
the free energy is often expressed in terms of the radius of a cluster,
$\widetilde{\Omega}\left(  R\right)  \equiv\Omega\left(  N\left(  R\right)
\right)  $ and one could equally well suppose that the distribution of cluster
radii is $\widetilde{P}\left(  R\right)  \sim\exp\left(  -\beta
\widetilde{\Omega}\left(  R\right)  \right)  $. However, these two expressions
cannot both be true because, strictly speaking, there is the exact relation
$P\left(  N\right)  dN=$ $\widetilde{P}\left(  R\right)  dR$ from which one
would deduce, e.g., $\widetilde{P}\left(  R\right)  =P\left(  N\left(
R\right)  \right)  \frac{dN}{dR}$or, more suggestively, $\widetilde{P}\left(
R\right)  \sim\exp\left(  -\beta\left(  \widetilde{\Omega}\left(  R\right)
+k_{B}T\ln\frac{dN\left(  R\right)  }{dR}\right)  \right)  $. Alternatively,
one could take $\widetilde{P}\left(  R\right)  $ as being more fundamental and
deduce $P\left(  N\right)  $ from it with a corresponding shift of the
effective free energy. Which of these is correct - or are they both wrong?\ In
the CNT limit, these distinctions are unimportant since $\ln\frac{dN\left(
R\right)  }{dR}\sim\ln R$ and so the correction to the free energy is
logarithmic and, hence, sub-dominant compared to the volume ($R^{D}$) and
surface ($R^{D-1}$) contributions that determine $\widetilde{\Omega}\left(
R\right)  $ in $D$ dimensions. However, in the description of small clusters,
even at low supersaturations, these distinctions become important. One way to
resolve this ambiguity would be to formulate a Fokker-Planck equation for the
probability density and then to use this to determine the stationary (e.g.
equilibrium) state. This is, in essence, the approach followed here.

Recently, there has been considerable interest in the possibility of extending
CNT in two different but related directions. The first, discussed in detail by
Prestipino et al\cite{Prestipino} concerns the correction of the capillary
model for the free energy used in CNT whereas the second, discussed e.g. by
Lechner et al\cite{Lechner}, concerns the choice of the reaction coordinate
used to characterize the nucleation pathway. In both cases, of course, the
quantity of most direct interest is the nucleation rate. For example, in the
former work it was pointed out that sub-dominant terms in the expansion of the
free energy as a function of cluster size make non-negligible contributions to
the nucleation rate for small clusters. The goal of the present paper is to
demonstrate a similar analysis which has the advantage that the entire
nucleation processes can be described self-consistently. This is important
because it turns out, as indicated above, that the choice of reaction
coordinate also involves sub-dominant corrections to the nucleation rate so
that the two avenues of investigation mentioned above are actually seen to
come together.

The present analysis is based on a recently developed description of
nucleation based on fluctuating
hydrodynamics\cite{Lutsko_JCP_2011_Com,Lutsko_JCP_2012_1,Lutsko_JCP_2012_2}.
Its range of validity is restricted to systems governed by a diffusive
dynamics typical of colloids or of macromolecules in solution. The new
formulation was motivated by difficulties in finding a consistent way to use
the tools of DFT\ (an equilibrium theory) to describe the process of
nucleation (a nonequilibrium process). In this approach, attention is focused
on the formation of a single cluster and the fundamental quantity is the
spatial density distribution that describes the cluster. One feature of the
theory is that the density distribution can be parametrized in terms of a few
physical quantities such as some measure of the size of a cluster, the width
of the interface, etc. and a dynamical description of the evolution of these
quantities results. When the cluster is parametrized by a single quantity, its
"size", one makes contact with CNT and, indeed, one can reproduce CNT in the
appropriate limits as described in Ref. \onlinecite{Lutsko_JCP_2012_1} and
below. (Subject to the same approximations such including that clusters are
spherical). In this paper, the goal is to extend this approach to allow for an
extension CNT that includes the finite width of the cluster interface in a
self-consistent manner. It proves relatively straightforward to extract a
nucleation rate from the formalism, essentially by following the development
from CNT, and a comparison between the predictions of the extended theory and
of CNT can be made. The ambiguity discussed above concerning the equilibrium
distribution is resolved and it is in fact found that neither the size nor the
radius are the most natural reaction coordinate.

In the next Section, the elements of Classical Nucleation Theory are reviewed
with a particular focus on the assumptions that go into it. The formulation
derived from fluctuating hydrodynamics is then described. It is noted that the
formulation possesses a type of covariance with respect to the choice of
reaction coordinate so that the ambiguities discussed above are resolved. An
expression for the nucleation rate is also derived. Section III shows how the
capillary model for a cluster can be used in conjunction with the dynamical
theory to reproduce familiar results from CNT such as the expression for the
monomer attachment rate, the nucleation rate and the rate of growth of
super-critical clusters. A model for clusters that allows for a finite
interfacial width is described in Section IV. Section V presents comparisons
between the models as well as tests of the assumptions underlying them. Our
results are summarized in the final Section.

\section{Theory}

\subsection{Classical Nucleation Theory}

In CNT, one begins by assuming that clusters of a given size can grow by the
addition of monomers, or can dissipate by spontaneously giving up a monomer to
the surrounding solution. Processes involving the interaction of clusters
which are both larger than monomers are ignored. Then, the concentration of
clusters of size $N$, $c_{N}$, is determined by a set of rate equations of the
form%
\begin{equation}
\frac{dc_{N}}{dt}=f_{N-1}c_{N-1}-g_{N}c_{N}+g_{N+1}c_{N+1}-f_{N}c_{N}
\label{A1}%
\end{equation}
where $f_{N}$ is the monomer attachment rate for a cluster of size $N$ and
$g_{N}$ is the monomer detachment rate for a cluster of size $N$. This is
known as the Becker-D\"{o}ring model\cite{Kashchiev,BeckerDoring}. For
condensation of a droplet from a diffuse solution, the monomer attachment rate
is clearly going to be proportional to the rate at which monomers impinge on
the cluster and so can be written as $f_{N}=\gamma_{N}j_{N}\left(  4\pi
R_{N}^{2}\right)  $ where $R_{N}$ is the radius of a cluster of size $N$,
$j_{N}$ is the rate per unit area at which molecules collide with the cluster
and $\gamma_{N}$ is a phenomenological factor describing the probability that
a colliding molecule actually sticks to the cluster. The rate at which
molecules in solution collide with the cluster can be determined by solving
the diffusion equation with adsorbing boundary conditions with the result that
in the long time (quasi-static approximation) $j_{N}=Dc_{1}/R_{N}$ where $D$
is the monomer diffusion constant, so that $f_{N}=\gamma_{N}\left(  4\pi
DR_{N}\right)  c_{1}$\cite{Kashchiev}. The monomer detachment rate is much
harder to estimate since it depends on the same physical details that are
accounted for phenomenologically in the sticking constant (i.e. details
concerning intermolecular interactions in the condensed phase).
Kashchiev\cite{Kashchiev} gives a general argument for a relation between the
monomer attachment frequency and the monomer detachment frequency for an
\emph{equilibrium} (i.e. under-saturated) system:\ namely, that in this case
detailed balance demands that $f_{N-1}c_{N-1}^{\left(  e\right)  }=g_{N}%
c_{N}^{\left(  e\right)  }$ so that, if we can assume a Boltzmann distribution
$c_{N}^{\left(  e\right)  }\propto\exp\left(  -\beta\Delta\Omega_{N}\right)  $
then it follows that $g_{N}=f_{N-1}\exp\left(  \beta\Delta\Omega_{N}%
-\beta\Delta\Omega_{N-1}\right)  $. Note that this argument only holds for the
under-saturated, equilibrium solution and that employing it for the
supersaturated, \emph{nonequilibrium} solution is a further assumption.

In the limit of large ( $N\gg1$) clusters, Eq.(\ref{A1}) can be approximated
by treating $c_{N}\left(  t\right)  \rightarrow C\left(  N;t\right)  $ as a
continuous function and expanding to get%
\begin{align}
\frac{dC\left(  N,t\right)  }{dt}  &  =\left(  f\left(  N\right)  C\left(
N;t\right)  -\epsilon\frac{\partial f\left(  N\right)  C\left(  N;t\right)
}{\partial N}+\frac{1}{2}\epsilon^{2}\frac{\partial^{2}f\left(  N\right)
C\left(  N;t\right)  }{\partial N^{2}}+...\right)  -g\left(  N\right)
C\left(  N;t\right) \\
&  +\left(  g\left(  N\right)  C\left(  N;t\right)  +\epsilon\frac{\partial
g\left(  N\right)  C\left(  N;t\right)  }{\partial N}+\frac{1}{2}\epsilon
^{2}\frac{\partial^{2}g\left(  N\right)  C\left(  N;t\right)  }{\partial
N^{2}}+...\right)  -f\left(  N\right)  C\left(  N;t\right) \nonumber
\end{align}
where a formal ordering parameter, $\epsilon$, has been introduced: it will,
at the end of the calculation, be set to one. Simplifying gives the Tunitskii
equation\cite{Kashchiev},%
\begin{equation}
\frac{dC\left(  N,t\right)  }{dt}=\frac{\partial}{\partial N}\left(
\epsilon\left(  g\left(  N\right)  -f\left(  N\right)  \right)  C\left(
N;t\right)  +\frac{1}{2}\epsilon^{2}\frac{\partial}{\partial N}\left(
f\left(  N\right)  +g\left(  N\right)  \right)  C\left(  N;t\right)  \right)
+O\left(  \epsilon^{3}\right)
\end{equation}
once we set $\epsilon=1$ and dropping the third and higher order
contributions. Similarly, expanding the approximation for the detachment rate
gives%
\begin{align}
g\left(  N\right)   &  =f\left(  N-1\right)  \exp\left(  \beta\Delta
\Omega\left(  N\right)  -\beta\Delta\Omega\left(  N-1\right)  \right)
\label{A2}\\
&  =f\left(  N\right)  -\epsilon\frac{\partial f\left(  N\right)  }{\partial
N}+\epsilon f\left(  N\right)  \frac{\partial\beta\Delta\Omega\left(
N\right)  }{\partial N}+O\left(  \epsilon^{2}\right) \nonumber
\end{align}
and combing with the Tunitskii equation results in
\begin{align}
\frac{dC\left(  N,t\right)  }{dt}  &  =\epsilon^{2}\frac{\partial}{\partial
N}\left(  \left(  -\frac{\partial f\left(  N\right)  }{\partial N}+f\left(
N\right)  \frac{\partial\beta\Delta\Omega\left(  N\right)  }{\partial
N}\right)  C\left(  N;t\right)  +\frac{\partial}{\partial N}f\left(  N\right)
C\left(  N;t\right)  \right)  +O\left(  \epsilon^{3}\right) \nonumber\\
&  =\epsilon^{2}\frac{\partial}{\partial N}\left(  f\left(  N\right)
\frac{\partial\beta\Delta\Omega\left(  N\right)  }{\partial N}C\left(
N;t\right)  +f\left(  N\right)  \frac{\partial}{\partial N}C\left(
N;t\right)  \right)  +O\left(  \epsilon^{3}\right)
\end{align}
which is the well-known Zeldovich equation\cite{Kashchiev}.

Although these manipulations are formally correct, there is in fact an
additional assumption being made. Although the derivative $\frac{\partial
f\left(  N\right)  }{\partial N}\sim\frac{f}{N}$ can be considered to be small
for large $N$, it is not clear that the same is true of the free energy since
in fact $\Delta\Omega\left(  N\right)  -\beta\Delta\Omega\left(  N-1\right)
\sim O\left(  1\right)  $. Taking only the first term in the expansion of the
exponential in Eq.(\ref{A2}) is therefore only justified if it can be assumed
that $\frac{\partial\beta\Delta\Omega\left(  N\right)  }{\partial N}\ll
1$:\ \emph{in other words, in the vicinity of the critical cluster}. It is
therefore clear that the Zeldovich equation is, strictly speaking, only
applicable for large $N$ and for $N$ near the critical size.

Noting that the probability to observe a cluster of size $N$, $P\left(
N\right)  \equiv\frac{n\left(  N\right)  }{\sum_{N^{\prime}}n\left(
N^{\prime}\right)  }$ where $n(N)$ is the number of clusters of size $N$, it
is easily seen that $C\left(  N;t\right)  =C_{tot}\left(  t\right)  P\left(
N;t\right)  $ where $C_{tot}\left(  t\right)  =\sum_{N}C\left(  N;t\right)  $
is the total number of molecules per unit volume, i.e. the average
concentration. So, the Zeldovich equation implies a Fokker-Planck equation for
$P(N;t)$%
\begin{equation}
\label{FP1}\frac{dP\left(  N,t\right)  }{dt}=\frac{\partial}{\partial
N}\left(  f\left(  N\right)  \frac{\partial\beta\Delta\Omega\left(  N\right)
}{\partial N}P\left(  N;t\right)  +f\left(  N\right)  \frac{\partial}{\partial
N}P\left(  N;t\right)  \right)  -P\left(  N;t\right)  \frac{\partial\ln
C_{tot}\left(  t\right)  }{\partial t}%
\end{equation}
where the source term vanishes if the total number of molecules is constant.

\subsection{Generalization based on fluctuating hydrodynamics}

In general, nucleation is a fluctuation-driven phenomena limited by the rate
of transport of mass, heat, etc. within the system. In fluid systems, it is
therefore naturally described by Landau's fluctuating hydrodynamics which
includes both transport and thermal fluctuations\cite{Landau, Chavanis_Deriv}.
The description of nucleation via fluctuating hydrodynamics has been described
in Refs. \onlinecite{Lutsko_JCP_2011_Com,Lutsko_JCP_2012_1,Lutsko_JCP_2012_2}.
A particularly simple description is possible when the system is diffusion
limited and over-damped, as is an appropriate description for colloids and
macromolecules in solution. The dynamics can be reduced to an equation for the
time-dependent spatial density and a reduced description is possible when the
density field is parametrized in the form%
\begin{equation}
\rho\left(  r;t\right)  \rightarrow\rho\left(  r;X_{1}\left(  t\right)
,X_{2}\left(  t\right)  ...X_{N}\left(  t\right)  \right)
\end{equation}
so that the time dependence occurs via a set of $N$ parameters which could
include, e.g., the size of the cluster, the width of the interface, the
density at the center of the cluster, etc. Here, we will be interested in the
case that there is a single parameter. Then, it was
shown\cite{Lutsko_JCP_2011_Com,Lutsko_JCP_2012_1,Lutsko_JCP_2012_2} that the
evolution of the order parameter may, in the case of an infinite system, be
approximated by a stochastic differential equation of the form%
\begin{equation}
\frac{dX}{dt}=-Dg^{-1}\left(  X\right)  \frac{\partial\beta\Omega}{\partial
X}-D\frac{1}{2}g^{-2}\left(  X\right)  \frac{\partial g\left(  X\right)
}{\partial X}+\sqrt{2Dg^{-1}\left(  X\right)  }\xi\left(  t\right)
\label{SDE}%
\end{equation}
where $D$ is the diffusion constant, $\Omega\left(  X\right)  =F\left(
x\right)  -\mu N\left(  X\right)  $, is the grand potential for chemical
potential $\mu$ and where $N\left(  X\right)  $ is the total number of
molecules. The last term on the right is proportional to $\xi\left(  t\right)
$ which is a delta-function correlated fluctuating force with mean zero and
variance equal to one. The quantity $g\left(  X\right)  $ that determines the
kinetic coefficient of the SDE as well as the amplitude of the noise is given
by%
\begin{equation}
g\left(  X\right)  =\int_{0}^{\infty}\frac{1}{4\pi r^{2}\rho\left(
r;X\right)  }\left(  \frac{\partial m\left(  r;X\right)  }{\partial X}\right)
^{2}dr \label{g1}%
\end{equation}
where the cumulative mass is%
\begin{equation}
m\left(  r;X\right)  =4\pi\int_{0}^{r}\rho\left(  r;X\right)  r^{\prime
2}dr^{\prime}.
\end{equation}
Since the amplitude of the noise in the SDE is state-dependent, it is
important to note that the equation is given according to the Ito
interpretation. The function $g(X)\ $will be called "the metric" as it can be
used to define a meaningful length in the general
theory\cite{Lutsko_JCP_2012_1} , although it will not be used for that purpose here.

\subsubsection{Fokker-Planck equation}

It is useful to also consider the equivalent Fokker-Planck equation which
is\cite{Gardiner}%
\begin{equation}
\frac{\partial P\left(  X,t\right)  }{\partial t}=D\frac{\partial}{\partial
X}\left(  g^{-1}\left(  X\right)  \frac{\partial\beta\Omega\left(  X\right)
}{\partial X}+g^{-1/2}\left(  X\right)  \frac{\partial}{\partial X}%
g^{-1/2}\left(  X\right)  \right)  P\left(  X,t\right)  \label{FPE1}%
\end{equation}
Note that this can be written as%
\begin{equation}
\frac{\partial P\left(  X,t\right)  }{\partial t}=D\frac{\partial}{\partial
X}\left(  g^{-1}\left(  X\right)  \frac{\partial\left(  \beta\Omega\left(
X\right)  -\ln g^{1/2}\left(  X\right)  \right)  }{\partial X}+g^{-1}\left(
X\right)  \frac{\partial}{\partial X}\right)  P\left(  X,t\right)
\label{FPE2}%
\end{equation}
so that comparison with Eq.(\ref{FP1}) shows that these are formally the same
with the monomer attachment frequency replaced by $g^{-1}\left(  X\right)  $
and with the free energy shifted by a term logarithmic in $g\left(  X\right)
$. From the expression for $g\left(  X\right)  $, Eq.(\ref{g1}), it is evident
that this function varies as the volume of the cluster so that the shift to
the free energy will go like the log of the radius of the cluster.

Let us look for a stationary solution determined, for some constant $J_{s}$ ,
from%
\begin{equation}
-D\left(  g^{-1}\left(  X\right)  \frac{\partial\beta\Omega\left(  X\right)
}{\partial X}+g^{-1/2}\left(  X\right)  \frac{\partial}{\partial X}%
g^{-1/2}\left(  X\right)  \right)  P_{s}\left(  X\right)  =J_{s},
\end{equation}
giving%
\begin{equation}
P_{s}\left(  X\right)  =Ag^{1/2}\left(  X\right)  \exp\left(  -\beta
\Omega\left(  X\right)  \right)  -D^{-1}J_{s}g^{1/2}\left(  X\right)
\exp\left(  -\beta\Omega\left(  X\right)  \right)  \int^{X}g^{1/2}\left(
X^{\prime}\right)  \exp\left(  \beta\Omega\left(  X^{\prime}\right)  \right)
dX^{\prime} \label{Ps}%
\end{equation}
for some constant $A$. Now, if we note that for some change of variables
$X\rightarrow Y\left(  X\right)  $ we have that%
\begin{equation}
\widetilde{g}\left(  Y\right)  =g\left(  X\left(  Y\right)  \right)  \left(
\frac{dX}{dY}\right)  ^{2}%
\end{equation}
and so%
\begin{align}
\widetilde{P}_{s}\left(  Y\right)   &  =A\widetilde{g}^{1/2}\left(  Y\right)
\exp\left(  -\beta\Omega\left(  Y\right)  \right)  -D^{-1}J_{s}\widetilde{g}%
^{1/2}\left(  Y\right)  \exp\left(  -\beta\Omega\left(  Y\right)  \right)
\int^{Y}\widetilde{g}^{1/2}\left(  Y^{\prime}\right)  \exp\left(  \beta
\Omega\left(  Y^{\prime}\right)  \right)  dY^{\prime}\\
&  =Ag^{1/2}\left(  X\right)  \frac{dX}{dY}\exp\left(  -\beta\Omega\left(
X\right)  \right) \nonumber\\
&  -D^{-1}J_{s}g^{1/2}\left(  X\right)  \frac{dX}{dY}\exp\left(  -\beta
\Omega\left(  X\right)  \right)  \int^{X}g^{1/2}\left(  X^{\prime}\right)
\frac{dX^{\prime}}{dY^{\prime}}\exp\left(  \beta\Omega\left(  X^{\prime
}\right)  \right)  dY^{\prime}\nonumber\\
&  =P_{s}\left(  X\right)  \frac{dX}{dY}\nonumber
\end{align}
Hence, this solution is completely covariant - there is not ambiguity as to
which variable is used. If the system is stable (i.e. under-saturated), then
it makes sense to seek an equilibrium ( $J=0$) solution which is
\begin{equation}
P_{eq}\left(  X\right)  =Ag^{1/2}\left(  X\right)  \exp\left(  -\beta
\Omega\left(  X\right)  \right)  =A\exp\left(  -\beta\left(  \Omega\left(
X\right)  -\frac{1}{2}k_{B}T\ln g\left(  X\right)  \right)  \right)
\end{equation}
where the constant $A$ is fixed by normalization. Note that this has the form
of a canonical distribution in terms of the shifted free energy.

\subsubsection{A Canonical Variable}

In the present, single-variable, case there is always a special variable for
which these expressions simplify. Given any variable, $X$, it is defined via%
\begin{equation}
dY=\sqrt{g\left(  X\right)  }dX
\end{equation}
and an arbitrary boundary condition that will be taken to be $Y\left(
0\right)  =0$. In terms of this variable, the Langevin equation becomes%
\begin{equation}
\frac{dY}{dt}=-D\frac{\partial\beta\widetilde{\Omega}\left(  Y\right)
}{\partial Y}+\sqrt{D}\xi\left(  t\right)
\end{equation}
where $\widetilde{\Omega}\left(  Y\right)  =\Omega\left(  X\left(  Y\right)
\right)  $. The corresponding Fokker-Planck equation is
\begin{equation}
\frac{\partial\widetilde{P}\left(  Y,t\right)  }{\partial t}=D\frac{\partial
}{\partial Y}\left(  \frac{\partial\beta\widetilde{\Omega}\left(  Y\right)
}{\partial Y}+\frac{\partial}{\partial Y}\right)  \widetilde{P}\left(
Y,t\right)  .
\end{equation}
These are equations that might have been written down on phenomenological
grounds but they would have been ambiguous since there is no a priori reason
to use the mass as the independent variable rather than, say, the equimolar
radius. This illustrates the way in which grounding the theory on a more
fundamental description serves to remove such ambiguities.

\subsubsection{Nucleation rate}

The standard argument to obtain the nucleation rate has a long history, as
discussed by Hanggi et al\cite{Hanggi}, and is based on boundary conditions
according to which (a)\ the distribution is stationary and (b) the
distribution goes to zero at some point beyond the critical cluster, say at
$X=X_{+}$(see also Refs. (\onlinecite{Kashchiev, Lifshitz})). The latter
condition represents the physical fact that once a cluster is slightly larger
than the critical cluster, it will almost certainly grow forever and so plays
no role in the stochastic part of the process. One therefore imagines
establishing a steady state by removing clusters once they reach size
$X=X_{+}$ and simultaneously re-injecting the removed material in the form of
monomers. Here, since we are only interested in the case of one-dimensional
barrier-crossing, this rate is easily evaluated from the exact solution for
the steady-state distribution. (For multivariate problems, further
approximations are necessary as described in Refs. \onlinecite{Langer1, Langer2, Hanggi}).

First, the nucleation rate, $J$, is the rate of production of super-critical
nuclei per unit volume and is therefore%
\begin{equation}
J=\frac{d}{dt}\int_{N^{\ast}}^{\infty}\frac{1}{V}n\left(  N;t\right)
dN=\frac{d}{dt}\int_{N^{\ast}}^{\infty}\frac{N_{c}\left(  t\right)  }%
{V}P\left(  N;t\right)  dN
\end{equation}
where $N_{c}\left(  t\right)  =\int_{0}^{\infty}n\left(  N;t\right)  dN$ is
the total number of clusters. Making use of the Fokker-Planck equation, this
becomes%
\begin{align}
J  &  =\frac{1}{V}\frac{dN_{c}\left(  t\right)  }{dt}\int_{N^{\ast}+1}%
^{\infty}P\left(  N,t\right)  dN\\
&  -D\frac{N_{c}\left(  t\right)  }{V}\left(  g^{-1}\left(  N\right)
\frac{\partial\beta\Omega\left(  N\right)  }{\partial N}P\left(  N,t\right)
+g^{-1/2}\left(  N\right)  \frac{\partial}{\partial N}g^{-1/2}\left(
N\right)  P\left(  N,t\right)  \right)  _{N=N^{\ast}}\nonumber
\end{align}
In the artificially imposed stationary state, the first term on the right does
not contribute leaving%
\begin{equation}
J=-D\frac{N_{c}}{V}\left(  g^{-1}\left(  N\right)  \frac{\partial\beta
\Omega\left(  N\right)  }{\partial N}P_{s}\left(  N\right)  +g^{-1/2}\left(
N\right)  \frac{\partial}{\partial N}g^{-1/2}\left(  N\right)  P_{s}\left(
N\right)  \right)  _{N=N^{\ast}}%
\end{equation}
Since we remove clusters of size $X_{+}$ as they form, we need a steady-state
distribution that satisfies $P\left(  X_{+}\right)  =0$. When this is imposed,
the general steady-state distribution, Eq.(\ref{Ps}), becomes%
\begin{equation}
P_{s}\left(  X\right)  =D^{-1}J_{s}g^{1/2}\left(  X\right)  \exp\left(
-\beta\Omega\left(  X\right)  \right)  \int_{X}^{X_{+}}g^{1/2}\left(
X^{\prime}\right)  \exp\left(  \beta\Omega\left(  X^{\prime}\right)  \right)
dX^{\prime} \label{stationary}%
\end{equation}
Substitution into the expression for the nucleation rate gives
\begin{equation}
J=\frac{N_{c}}{V}J_{s}%
\end{equation}
The remainder of the development concerns the relation between the imposed
stationary flux, $J_{s}$, and real properties of the system. Zeldovich, as
described by Kashchiev\cite{Kashchiev} requires that the concentration of
monomers be fixed at the "equilibrium" value%
\begin{equation}
\rho_{\infty}=\frac{N_{c}}{V}P_{s}\left(  X_{1}\right)  \label{Zc}%
\end{equation}
where $X_{1}$ is the value of the parameter $X$ that fixes the number of
molecules in the cluster to be $1$. (The notation chosen here is motivated by
the fact that one expects that $\rho_{infty}$ is the density far from a
cluster.) This then gives%
\begin{align}
J_{CNT}  &  =\frac{D\rho_{\infty}}{g^{1/2}\left(  X_{1}\right)  \exp\left(
-\beta\Omega\left(  X_{1}\right)  \right)  \int_{X_{1}}^{X_{+}}g^{1/2}\left(
X^{\prime}\right)  \exp\left(  \beta\Omega\left(  X^{\prime}\right)  \right)
dX^{\prime}}\label{Jcnt}\\
&  =\frac{D\rho_{\infty}}{\int_{X_{1}}^{X_{+}}g\left(  X^{\prime}\right)
\exp\left(  \Delta\beta\Omega\left(  X^{\prime}\right)  -\frac{1}{2}\ln
\frac{g\left(  X^{\prime}\right)  }{g\left(  X_{1}\right)  }\right)
dX^{\prime}}\nonumber
\end{align}
The second equality gives the result written in terms of the adjusted free
energy. A saddle-point evaluation for the choice of number of molecules as the
variable gives%
\begin{equation}
J_{CNT}\simeq D\rho_{\infty}g^{-1}\left(  N_{\ast\ast}\right)  \sqrt{\frac
{1}{2\pi}}\sqrt{\left\vert \Delta\beta\Omega^{\prime\prime}\left(  N_{\ast
\ast}\right)  \right\vert }\exp\left(  -\Delta\beta\Omega_{\ast\ast}\right)
\label{Jcnt-ap}%
\end{equation}
where the adjusted critical size is determined from
\begin{equation}
\Delta\beta\Omega^{\prime}\left(  N_{\ast\ast}\right)  -\frac{1}{2}%
\frac{g^{\prime}\left(  N_{\ast\ast}\right)  }{g\left(  N_{\ast\ast}\right)
}=0
\end{equation}
and where
\begin{equation}
\Delta\beta\Omega_{\ast\ast}\equiv\Delta\beta\Omega\left(  N_{\ast\ast
}\right)  -\frac{1}{2}\ln\frac{g\left(  N_{\ast\ast}\right)  }{g\left(
X_{1}\right)  }%
\end{equation}
The problem with this expression is that it is not covariant and the result
will depend on the choice of $X$. This is due to the fact that the condition,
fixing the fraction of monomers, is itself not covariant since only the
combination $P_{s}\left(  X\right)  dX$ is invariant under a change of
variables. We stress that within the context of CNT this fact is irrelevant
since the assumption of large $N$ means that the lack of covariance is due to
sub-dominant terms in the distribution. However, within the context of the
general theory, this lack of covariance indicates unnecessary ambiguity since
different results will be obtained for different choices of the variable $X$
appearing in the boundary condition, Eq.( \ref{Zc}).

The first step to solving this problem is to note that, as defined here,
$P_{s}\left(  X\right)  $ is normalized so that we can immediately fix the
unknown coefficient giving%
\begin{equation}
P_{s}\left(  X\right)  =\frac{g^{1/2}\left(  X\right)  \exp\left(
-\beta\Omega\left(  X\right)  \right)  \int_{X}^{X_{+}}g^{1/2}\left(
X^{\prime}\right)  \exp\left(  \beta\Omega\left(  X^{\prime}\right)  \right)
dX^{\prime}}{\int_{0}^{X_{+}}dX\;g^{1/2}\left(  X\right)  \exp\left(
-\beta\Omega\left(  X\right)  \right)  \int_{X}^{X_{+}}g^{1/2}\left(
X^{\prime}\right)  \exp\left(  \beta\Omega\left(  X^{\prime}\right)  \right)
dX^{\prime}}%
\end{equation}
and%
\begin{equation}
J=\frac{N_{c}}{V}\frac{D}{\int_{0}^{X_{+}}dX\;g^{1/2}\left(  X\right)
\exp\left(  -\beta\Omega\left(  X\right)  \right)  \int_{X}^{X_{+}}%
g^{1/2}\left(  X^{\prime}\right)  \exp\left(  \beta\Omega\left(  X^{\prime
}\right)  \right)  dX^{\prime}}%
\end{equation}
This still leaves the question of determining the overall concentration of
clusters. We do this following the general idea used in CNT but being careful
to preserve covariance by imposing a condition on the total number of
molecules, $\mathcal{N}\left(  N_{0}\right)  $ in clusters up to size $N_{0}$,%
\begin{equation}
\mathcal{N}\left(  N_{0}\right)  =\int_{0}^{N_{0}}Nn\left(  N\right)
dN=N_{c}\int_{0}^{N_{0}}NP_{s}\left(  N\right)  dN
\end{equation}
so that%
\begin{equation}
N_{c}=\mathcal{N}\left(  N_{0}\right)  \frac{\int_{0}^{X_{+}}dXg^{1/2}\left(
X\right)  \exp\left(  -\beta\Omega\left(  X\right)  \right)  \int_{X}^{X_{+}%
}g^{1/2}\left(  X^{\prime}\right)  \exp\left(  \beta\Omega\left(  X^{\prime
}\right)  \right)  dX^{\prime}}{\int_{0}^{X\left(  N_{0}\right)  }dX\;N\left(
X\right)  g^{1/2}\left(  X\right)  \exp\left(  -\beta\Omega\left(  X\right)
\right)  \int_{X}^{X_{+}}g^{1/2}\left(  X^{\prime}\right)  \exp\left(
\beta\Omega\left(  X^{\prime}\right)  \right)  dX^{\prime}}%
\end{equation}
giving the nucleation rate as%
\begin{equation}
J\left(  N_{0}\right)  =\frac{D\overline{\rho}\left(  N_{0}\right)  }{\int%
_{0}^{X\left(  N_{0}\right)  }dX\;N\left(  X\right)  g^{1/2}\left(  X\right)
\exp\left(  -\beta\Omega\left(  X\right)  \right)  \int_{X}^{X\left(
N_{+}\right)  }g^{1/2}\left(  X^{\prime}\right)  \exp\left(  \beta
\Omega\left(  X^{\prime}\right)  \right)  dX^{\prime}} \label{Jexact}%
\end{equation}
where the average density is $\overline{\rho}\left(  N_{0}\right)
=\frac{\mathcal{N}\left(  N_{0}\right)  }{V}$ and where we have indicated that
the boundary condition makes the nucleation rate a function of $N_{0}$. If
$N_{0}$ is chosen to be large, say the size of the critical cluster $N_{\ast}%
$, then $\mathcal{N}\left(  N_{\ast}\right)  $ is the total population in the
initial (i.e. precritical) state and this is essentially the standard
expression for the reaction rate derived from one-dimensional barrier crossing
(the "flux over population" expression, see, e.g., Hanggi et al\cite{Hanggi})
except that an extra factor of $N$ occurs in the integral in the denominator,
which is meant to count the population in the initial state, because each
cluster contains many molecules. If $N_{0}$ is chosen to be on the order of
one, then we are essentially counting the number of monomers. Importantly, if
most of the material exists in the form of small clusters (monomers, dimers,
etc.) then these estimates will be the same. If they are not the same, then a
significant amount of mass is present in the form of clusters and the implicit
assumption (made here as well as in CNT) that clusters do not interact is
probably invalid. One of the goals below will therefore be to monitor the
validity of this assumption.

Assuming that the bulk of the material is in the form of small clusters and
that $N_{0}$ is chosen sufficiently large , an approximation to the exact
expression, Eq.(\ref{Jexact}), can be developed as described in Appendix
\ref{AppRate}. Assuming that the free energy and number of molecules as
functions of the canonical variable $Y$ have the expansions $\Delta
\beta\widetilde{\Omega}\left(  Y\right)  =\widetilde{\Omega}_{0}Y^{\alpha
}+...$ and $N\left(  Y\right)  =n_{0}Y^{\beta}+...$ for small $Y$, it turns
out that the approximation implies that, for small $X$, the stationary
distribution can be approximated by
\begin{equation}
P_{s}\left(  X\right)  \sim\frac{\alpha\widetilde{\Omega}_{0}^{1/\alpha}%
}{\Gamma\left(  \frac{1}{\alpha}\right)  }g^{1/2}\left(  X\right)  \exp\left(
-\Delta\beta\Omega\left(  X\right)  \right)  \label{PsApp}%
\end{equation}
and in which case the nucleation rate is approximated by%
\begin{equation}
J \sim\frac{1}{\sqrt{2\pi}}\rho_{\infty}D\frac{\alpha\widetilde{\Omega}%
_{0}^{\frac{\beta+1}{\alpha}}}{\Gamma\left(  \frac{\beta+1}{\alpha}\right)
n_{0}}\sqrt{\left\vert \beta\widetilde{\Omega}^{\prime\prime}\left(  X_{\ast
}\right)  g^{-1}\left(  X_{\ast}\right)  \right\vert }\exp\left(  -\Delta
\beta\Omega_{\ast}\right)  \label{Japp}%
\end{equation}

\section{Capillary model: Classical nucleation theory}

The process we will describe is the nucleation of a liquid, with bulk density
$\rho_{l}$, from a vapor with bulk density $\rho_{v}$ at some temperature $T$
and chemical potential $\mu$. Let $\omega\left(  \rho\right)  $ be the free
energy (grand potential) per unit volume, so that $\omega\left(  \rho\right)
=f\left(  \rho\right)  -\mu\rho$ where $f\left(  \rho\right)  $ is the
Helmholtz free energy per unit volume. Then, the liquid and vapor densities
are determined by the imposed chemical potential via $\omega^{\prime}\left(
\rho_{l}\right)  =\omega^{\prime}\left(  \rho_{v}\right)  =0$ and since we
choose thermodynamic conditions such that the vapor is metastable,
$\omega\left(  \rho_{l}\right)  <\omega\left(  \rho_{v}\right)  $. To recover
CNT it is only necessary to use the capillary model for the density%
\begin{equation}
\rho\left(  r\right)  =\left\{
\begin{array}
[c]{c}%
\rho_{0},\;\;r<R\\
\rho_{\infty},\;\;R<r
\end{array}
\right.
\end{equation}
where we take $\rho_{0}=\rho_{l}$ for the density inside the cluster and
$\rho_{\infty}=\rho_{v}$ for the density outside the cluster. The
capillary-theory expression for the free energy of the cluster is%
\begin{equation}
\Delta\beta\Omega\left(  R\right)  =\frac{4\pi}{3}R^{3}\Delta\beta
\omega\left(  \rho_{0}\right)  +4\pi R^{2}\gamma\label{cap1}%
\end{equation}
where the second term represents the effect of surface tension. Note that the
only parameter that is allowed to vary is the radius of the cluster.

The model for the density gives the cumulative mass density%
\begin{equation}
m\left(  r\right)  =\left\{
\begin{array}
[c]{c}%
\frac{4\pi}{3}r^{3}\rho_{0},\;\;r<R\\
\frac{4\pi}{3}R^{3}\rho_{0}+\frac{4\pi}{3}\left(  r^{3}-R^{3}\right)
\rho_{\infty},\;\;R<r
\end{array}
\right.
\end{equation}
and the metric%
\begin{equation}
g\left(  R\right)  =\int_{R}^{\infty}\frac{1}{4\pi r^{2}\rho_{\infty}}\left(
4\pi R^{2}\left(  \rho_{0}-\rho_{\infty}\right)  \right)  ^{2}dr=\frac{\left(
\rho_{0}-\rho_{\infty}\right)  ^{2}}{\rho_{\infty}}4\pi R^{3}%
\end{equation}
The excess number of molecules in the cluster is
\begin{equation}
\Delta N=\frac{4\pi}{3}R^{3}\left(  \rho_{0}-\rho_{\infty}\right)
\end{equation}
and the metric in terms of the number of molecules is
\begin{equation}
g\left(  \Delta N\right)  =g\left(  R\right)  \left(  \frac{dR}{d\Delta
N}\right)  ^{2}=\frac{1}{4\pi\rho_{\infty}R\left(  \Delta N\right)  }.
\end{equation}
The canonical variable is
\begin{equation}
Y=\int_{0}^{R}\sqrt{\frac{\left(  \rho_{0}-\rho_{\infty}\right)  ^{2}}%
{\rho_{\infty}}4\pi R^{\prime3}}dR^{\prime}=\frac{2}{5}\sqrt{4\pi\frac{\left(
\rho_{0}-\rho_{\infty}\right)  ^{2}}{\rho_{\infty}}}R^{5/2}.
\end{equation}

In comparing the Fokker-Planck equation for the generalized theory,
Eq.(\ref{FPE2}), to that of Zubarev, Eq(\ref{FP1}), one finds that agreement
provided that the logarithmic corrections to the free energy are neglected and
the effective monomer attachment frequency is identified as
\begin{equation}
f\left(  N\right)  =g^{-1}\left(  \Delta N\right)  =4\pi\rho_{\infty}R\left(
\Delta N\right)
\end{equation}
which is the usual result for diffusion-limited nucleation in the case that
the phenomenological "sticking constant"\ is equal to one\cite{Kashchiev}. The
stochastic differential equation for the radius is%
\begin{equation}
\frac{dR}{dt}=-D\frac{\rho_{\infty}}{\left(  \rho_{0}-\rho_{\infty}\right)
^{2}4\pi R^{3}}\frac{\partial\left(  \beta\Omega-\ln g^{1/2}\left(  R\right)
\right)  }{\partial R}+\sqrt{2Dg^{-1}\left(  R\right)  }\xi\left(  t\right)
\end{equation}
When the cluster is large (i.e. when it is super-critical) the noise becomes
unimportant and the radius grows as
\begin{equation}
\frac{dR}{dt}=\frac{D\rho_{\infty}\left\vert \Delta\beta\omega\left(  \rho
_{0}\right)  \right\vert }{\left(  \rho_{0}-\rho_{\infty}\right)  ^{2}}%
R^{-1}+O\left(  R^{-2}\right)
\end{equation}
which gives the classical result $R\sim t^{1/2}$when the higher order terms
are neglected\cite{Saito}. In the weak liquid limit, $\rho_{\infty}\ll\rho
_{0}$, the coefficient of $R^{-1}$ agrees with that given by Lifshitz et
al\cite{Lifshitz}.

The (non-covariant) CNT-like nucleation rate, from\ Eq.(\ref{Jcnt-ap}), is%
\begin{equation}
J_{CNT}\sim2\sqrt{2\pi}D\rho_{\infty}R\left(  \Delta N_{\ast\ast}\right)
\sqrt{\left\vert \Delta\beta\Omega^{\prime\prime}\left(  \Delta N_{\ast\ast
}\right)  \right\vert }\exp\left(  -\Delta\beta\Omega_{\ast\ast}\right)
\end{equation}
and in the limit that the logarithmic corrections to the free energy are
negligible, this agrees with the result from\ CNT. In the following, we will
use as a reference the usual CNT result that is obtained by ignoring the
logarithmic shift in the free energy,%
\begin{align}
J_{CNT}  &  \sim2\sqrt{2\pi}D\rho_{\infty}R\left(  \Delta N_{\ast}\right)
\sqrt{\left\vert \Delta\beta\Omega^{\prime\prime}\left(  \Delta N_{\ast
}\right)  \right\vert }\exp\left(  -\Delta\beta\Omega_{\ast}^{CNT}\right) \\
&  =\frac{D\rho_{\infty}^{2}}{\rho_{0}}\frac{\left\vert \Delta\beta
\omega\right\vert }{\sqrt{\gamma}}\exp\left(  -\Delta\beta\Omega_{\ast}%
^{CNT}\right)  ,\;\Delta\beta\Omega_{\ast}^{CNT}=\frac{16\pi}{3}\frac
{\gamma^{3}}{\left(  \Delta\beta\omega\right)  ^{2}}.\nonumber
\end{align}
The approximate nucleation rate based on a condition of fixed mass,
Eq(\ref{Japp}), is
\begin{align}
\frac{J}{J_{CNT}}  &  \sim\frac{3\rho_{0}\sqrt{2}\left(  4\pi\right)
^{\frac{3}{4}}}{\Gamma\left(  \frac{11}{4}\right)  \sqrt{\rho_{\infty}%
}\left\vert \Delta\beta\omega\right\vert \left(  \rho_{0}-\rho_{\infty
}\right)  ^{2}}\gamma^{\frac{13}{4}}\left\vert \beta\Omega^{\prime\prime
}\left(  R_{\ast}\right)  g^{-1}\left(  R_{\ast}\right)  \right\vert
^{1/2}\label{kcnt}\\
&  =\frac{3\sqrt{2}\left(  4\pi\right)  ^{\frac{3}{4}}}{\Gamma\left(
\frac{11}{4}\right)  }\frac{\rho_{0}}{\left(  \rho_{0}-\rho_{\infty}\right)
^{3}}\gamma^{\frac{9}{4}}\left\vert \Delta\beta\omega\left(  \rho_{0}\right)
\right\vert ^{1/2}.\nonumber
\end{align}
Note that in all cases, we are assuming that the average concentration of
material in the (hypothetical) steady state, $\overline{\rho}$, is the same as
the background density, $\rho_{\infty}$.

\section{Extended Model: Finite cluster width}

\subsection{The cluster structure and the metric}

A significant short coming of the capillary model is that the width of the
cluster's interface is zero. A more realistic model will have a finite width
which, from various simulations and DFT calculations, might be expected to be
two or three molecular diameters in width. A simple extension of the capillary
model to take account of a finite width is the piecewise-linear profile,%
\begin{equation}
\rho\left(  r\right)  =\left\{
\begin{array}
[c]{c}%
\rho_{0},\;\;r<R-w\\
\rho_{0}-\left(  \rho_{0}-\rho_{\infty}\right)  \frac{r-\left(  R-w\right)
}{w},\;\;R-w<r<R\\
\rho_{\infty},\;\;R<r
\end{array}
\right.
\end{equation}
The corresponding cumulative mass distribution for $0\leq R\leq w$ is%
\begin{equation}
m\left(  r\right)  =\Theta\left(  R-r\right)  \frac{\pi}{3w}\left(  \rho
_{0}-\rho_{\infty}\right)  r^{3}\left(  4R-3r\right)  +\Theta\left(
r-R\right)  \frac{\pi}{3w}\left(  \rho_{0}-\rho_{\infty}\right)
R^{4}+V\left(  r\right)  \rho_{\infty}%
\end{equation}
while for $R>w$ it becomes%
\begin{align}
m\left(  r\right)   &  =\left(  \rho_{0}-\rho_{\infty}\right)  V\left(
r\right)  \Theta\left(  R-w-r\right) \\
&  +\Theta\left(  r-\left(  R-w\right)  \right)  \Theta\left(  R-r\right)
\frac{\pi}{3w}\left(  \rho_{0}-\rho_{\infty}\right)  \left(  r^{3}\left(
4R-3r\right)  -\left(  R-w\right)  ^{4}\right)  \allowbreak\allowbreak
\allowbreak\nonumber\\
&  +\Theta\left(  r-R\right)  \frac{\pi}{3w}\left(  \rho_{0}-\rho_{\infty
}\right)  \left(  R^{4}-\left(  R-w\right)  ^{4}\right) \nonumber\\
&  +V\left(  r\right)  \rho_{\infty}.\nonumber
\end{align}
Calculation of the metric is then straightforward with the result that for
$0\leq R\leq w$%
\begin{equation}
g\left(  R\right)  =4\pi\left(  \frac{\rho_{0}-\rho_{\infty}}{3w}\right)
^{2}\left[  \frac{wR^{4}}{\left(  \rho_{0}-\rho_{\infty}\right)  }\left[
a^{4}\ln\left(  \frac{a}{a-1}\right)  -\left(  a^{3}+\frac{1}{2}a^{2}+\frac
{1}{3}a+\frac{1}{4}\right)  \right]  +\frac{1}{\rho_{\infty}}R^{5}\right]
\end{equation}
with%
\begin{equation}
a=1+\frac{w\rho_{\infty}}{R\left(  \rho_{0}-\rho_{\infty}\right)  }.
\end{equation}
Note that for small clusters, $R\ll w$, this gives%
\begin{equation}
g\left(  R\right)  =\frac{8\pi}{15}\left(  \frac{\rho_{0}-\rho_{\infty}}%
{w}\right)  ^{2}\frac{R^{5}}{\rho_{\infty}}\left[  1-\frac{\left(  \rho
_{0}-\rho_{\infty}\right)  }{36\rho_{\infty}}\frac{R}{w}+O\left(  \left(
\frac{R}{w}\right)  ^{2}\right)  \right]
\end{equation}
For larger radii, $R>w$, the result is%
\begin{align}
&  g\left(  R\right)  =\frac{4\pi}{9}\frac{\left(  \rho_{0}-\rho_{\infty
}\right)  ^{2}}{\rho_{\infty}}\frac{\left(  R^{3}-\left(  R-w\right)
^{3}\right)  ^{2}}{w^{2}R}\\
&  +\frac{4\pi}{9}\left(  \frac{\rho_{0}-\rho_{\infty}}{w}\right)  \left(
\frac{\left(  R-w\right)  ^{6}}{R^{2}a^{2}}\ln\frac{R\allowbreak}{R-w}%
-\frac{\left(  \left(  R-w\right)  ^{3}-R^{3}a^{3}\right)  ^{2}}{R^{2}a^{2}%
}\ln\left(  \frac{\rho_{\infty}}{\rho_{0}}\right)  \right) \nonumber\\
&  -\frac{4\pi}{9}\left(  \frac{\rho_{0}-\rho_{\infty}}{w}\right)  \left(
\begin{array}
[c]{c}%
-w\frac{\left(  R-w\right)  ^{5}}{R^{2}a}-\left(  2\left(  R-w\right)
^{3}-R^{3}a^{3}\right)  w\\
+\frac{1}{2}R^{2}a^{2}\left(  R^{2}-\left(  R-w\right)  ^{2}\right)  +\frac
{1}{3}Ra\left(  R^{3}-\left(  R-w\right)  ^{3}\right)  +\frac{1}{4}\left(
R^{4}-\left(  R-w\right)  ^{4}\right)  \allowbreak\allowbreak
\end{array}
\right) \nonumber
\end{align}
One can again define a canonical variable using
\begin{equation}
\frac{dY}{dR}=\sqrt{g\left(  R\right)  }%
\end{equation}
and for small clusters one finds%
\begin{equation}
Y=\frac{2}{7}\left(  \frac{\rho_{0}-\rho_{\infty}}{w}\right)  \sqrt{\frac
{8\pi}{15\rho_{\infty}}}R^{\frac{7}{2}}\left(  1+O\left(  R\right)  \right)  .
\label{Yextended}%
\end{equation}

\subsection{Free Energy model}

It would be somewhat inconsistent to use the capillary approximation for the
free energy given that the assumed profile now has finite width. We therefore
consider a simple, but more fundamental free energy model based on the
squared-gradient approximation,%
\begin{equation}
\Omega\left[  \rho\right]  =\int\left(  \omega\left(  \rho\left(
\mathbf{r}\right)  \right)  +\frac{1}{2}K\left(  \mathbf{\nabla}\rho\left(
\mathbf{r}\right)  \right)  ^{2}\right)  d\mathbf{r}%
\end{equation}
where $\omega\left(  \rho\right)  =f\left(  \rho\right)  -\mu\rho$ is the
grand potential per unit volume, $f\left(  \rho\right)  $ is the Helmholtz
free energy per unit volume for a bulk system with uniform density $\rho$
which can be determined based on a given pair potential using thermodynamic
perturbation theory or liquid state integral equation methods. The
squared-gradient coefficient, $K$, can be estimated from a model interaction
potential using the results of Ref.[\onlinecite{Lutsko2011a}]. For the assumed
density profile, this becomes
\begin{align}
\Omega\left(  R;w\right)  -\Omega_{\infty}  &  =\frac{4\pi}{3}\left(
\max\left(  R-w,0\right)  \right)  ^{3}\Delta\omega\left(  \rho_{0}\right)
\label{a1}\\
&  +4\pi\int_{\max\left(  0,R-w\right)  }^{R}\Delta\omega\left(  \rho
_{0}-\left(  \rho_{0}-\rho_{\infty}\right)  \frac{r-\left(  R-w\right)  }%
{w}\right)  r^{2}dr\nonumber\\
&  +\frac{1}{2}K\frac{4\pi}{3}\left(  R^{3}-\max\left(  R-w,0\right)
^{3}\right)  \left(  \frac{\rho_{0}-\rho_{\infty}}{w}\right)  ^{2}\nonumber
\end{align}
The result for $R>w$, can also be written as
\begin{align}
\Omega\left(  R;w\right)  -\Omega_{\infty}  &  =\frac{4\pi}{3}\left(
R-w\right)  ^{3}\Delta\omega\left(  \rho_{0}\right) \label{b1a}\\
&  +4\pi\left(  2\overline{\omega}_{0}w+K\frac{\left(  \rho_{0}-\rho_{\infty
}\right)  ^{2}}{2w}\right)  R^{2}\nonumber\\
&  -4\pi\left(  2\overline{\omega}_{1}w+K\frac{\left(  \rho_{0}-\rho_{\infty
}\right)  ^{2}}{2w}\right)  Rw\nonumber\\
&  +4\pi\left(  \overline{\omega}_{2}w+K\frac{\left(  \rho_{0}-\rho_{\infty
}\right)  ^{2}}{6w}\right)  w^{2}\nonumber
\end{align}
where the density moments of the excess free energy per unit volume are%
\begin{equation}
\overline{\omega}_{n}=\frac{1}{\left(  \rho_{0}-\rho_{\infty}\right)  ^{n+1}%
}\int_{\rho_{\infty}}^{\rho_{0}}\left(  \omega\left(  x\right)  -\omega\left(
\rho_{\infty}\right)  \right)  \left(  x-\rho_{\infty}\right)  ^{n}dx,
\label{b2}%
\end{equation}
thus showing how, for large clusters (in the $\frac{w}{R}\rightarrow0$ limit),
one recovers something like the capillary approximation but with a variable
width. Minimizing with respect to the width at constant radius and solving as
an expansion in the radius (i.e. assuming $R\gg w$) gives (see Appendix
\ref{Expansion})%
\begin{equation}
w_{\min}=w_{0}\left(  1+\frac{\Delta\omega\left(  \rho_{0}\right)
-2\overline{\omega}_{1}}{\Delta\omega\left(  \rho_{0}\right)  -\overline
{\omega}_{0}}\frac{w_{0}}{R}+...\right)  \label{b3}%
\end{equation}
with%
\begin{equation}
w_{0}=\sqrt{\frac{K\left(  \rho_{0}-\rho_{\infty}\right)  ^{2}}{2\left(
\overline{\omega}_{0}-\Delta\omega\left(  \rho_{0}\right)  \right)  }}
\label{width}%
\end{equation}
and the free energy becomes
\begin{equation}
\Omega\left(  R;w\right)  -\Omega_{\infty}=\allowbreak\left(  \frac{4}{3}\pi
R^{3}\right)  \Delta\omega\left(  \rho_{0}\right)  +\left(  4\pi R^{2}\right)
\left(  \frac{1}{2}\Delta\omega\left(  \rho_{0}\right)  w_{0}+\frac{K\left(
\rho_{0}-\rho_{\infty}\right)  ^{2}}{w_{0}}\right)  +O\left(  R^{1}\right)
\label{b4}%
\end{equation}
The higher order terms are a simple illustration of the post-CNT corrections
to the free energy barrier recently discussed by Prestipino, et
al\cite{Prestipino}. At lowest order, the implied capillary-like model is
\begin{equation}
\Omega\left(  R;w\right)  -\Omega_{\infty}=\left(  \frac{4\pi}{3}R^{3}\right)
\Delta\omega\left(  \rho_{0}\right)  +\left(  \allowbreak4\pi R^{2}\right)
\gamma
\end{equation}
with%
\begin{align}
\gamma &  =\frac{1}{2}\Delta\omega\left(  \rho_{0}\right)  w_{0}%
+\frac{K\left(  \rho_{0}-\rho_{\infty}\right)  ^{2}}{w_{0}}\label{b5}\\
&  =\left(  1+\frac{\Delta\omega\left(  \rho_{0}\right)  }{4\left(
\overline{\omega}_{0}-\Delta\omega\left(  \rho_{0}\right)  \right)  }\right)
\sqrt{2K\left(  \rho_{0}-\rho_{\infty}\right)  ^{2}\left(  \overline{\omega
}_{0}-\Delta\omega\left(  \rho_{0}\right)  \right)  }.\nonumber
\end{align}
Note that the coefficient $\gamma$ depends on the supersaturation implicitly
via its dependence on the densities. In CNT, it is more common to ignore the
state dependence of this coefficient and to fix its value to that which gives
the correct value of the planar surface tension at coexistence. Here,
\textquotedblleft correct\textquotedblright\ will be taken to mean that it
gives the same value as the full model with finite width. Since at coexistence
the free energies of the two phases are equal, this is
\begin{equation}
\gamma_{CNT}=\left(  \rho_{0}^{coex}-\rho_{\infty}^{coex}\right)
\sqrt{2K\overline{\omega}_{0}^{coex}}. \label{gan-fixed}%
\end{equation}
where the superscripts indicate that the densities are those at coexistence.

One peculiarity of this model is that the value of the width that minimizes
the free energy undergoes a bifurcation as the radius increases. This is due
to the fact that for $w<R$ the free energy is%
\begin{equation}
\Omega\left(  R;w\right)  -\Omega_{\infty}=4\pi\int_{0}^{R}\Delta\omega\left(
\rho_{0}-\left(  \rho_{0}-\rho_{\infty}\right)  \frac{r-\left(  R-w\right)
}{w}\right)  r^{2}dr+\frac{2\pi}{3}KR^{3}\left(  \frac{\rho_{0}-\rho_{\infty}%
}{w}\right)  ^{2}%
\end{equation}
and it is clear the free energy difference decreases monotonically to zero as
the width increases.

Expanding in $R$ one has that%
\begin{equation}
\Omega\left(  R;w\right)  -\Omega_{\infty}=\frac{2\pi}{3}K\left(  \frac
{\rho_{0}-\rho_{\infty}}{w}\right)  ^{2}R^{3}+\frac{8\pi}{5!}\Delta
\omega^{\prime\prime}\left(  \rho_{\infty}\right)  \left(  \frac{\rho_{0}%
-\rho_{\infty}}{w}\right)  ^{2}R^{5}+O\left(  R^{6}\right)
\end{equation}
and in terms of the canonical variable%
\begin{equation}
\Omega\left(  Y;w\right)  -\Omega_{\infty}=\frac{2\pi}{3}K\left(  \frac
{\rho_{0}-\rho_{\infty}}{w}\right)  ^{2}\left(  \frac{7}{2}\left(  \frac
{w}{\rho_{0}-\rho_{\infty}}\right)  \sqrt{\frac{15\rho_{\infty}}{8\pi}%
}\right)  ^{\frac{6}{7}}Y^{\frac{6}{7}}+O\left(  Y^{\frac{10}{7}}\right)
\end{equation}
which can be used to evaluate the approximate nucleation rate%

\begin{equation}
J_{ext}\sim Z_{ext}\left\vert \beta\Omega^{\prime\prime}\left(  R_{\ast
}\right)  \right\vert ^{1/2}\exp\left(  -\beta\Omega_{\ast}\right)
\label{kext}%
\end{equation}
with the prefactor
\begin{equation}
Z_{ext}=D\frac{\sqrt{5}}{\Gamma\left(  \frac{13}{6}\right)  }\left(
\frac{9\pi}{2}\right)  ^{\frac{1}{6}}\sqrt{K\rho_{\infty}}\left(
\overline{\omega}_{0}-\Delta\omega\left(  \rho_{0}\right)  \right)  ^{\frac
{5}{3}}g^{-1/2}\left(  R_{\ast}\right)
\end{equation}

\section{Results and comparisons}

In order to illustrate the theory developed above, we have performed detailed
calculations for a model globular protein. We assume that the solvent can be
approximated, crudely, by assuming Brownian dynamics of the (large) solute
molecules which also experience an effective pair interaction for which we use
the ten Wolde-Frenkel interaction potential%
\begin{equation}
v\left(  r\right)  =\left\{
\begin{array}
[c]{c}%
\infty,\;\;r\leq\sigma\\
\,\frac{4\,\epsilon}{\alpha^{2}}\left(  \,\left(  \frac{1}{(\frac{r}{\sigma
})^{2}-1}\right)  ^{6}-\,\alpha\,\left(  \frac{1}{(\frac{r}{\sigma})^{2}%
-1}\right)  ^{3}\right)  ,\;\;r\geq\sigma
\end{array}
\right.
\end{equation}
with $\alpha=50$ which is then cutoff at $r_{c}=2.5\sigma$ and shifted so that
$v\left(  r_{c}\right)  =0$. The temperature is fixed at $k_{B}T=0.375\epsilon
$ and the equation of state computed using thermodynamic perturbation theory.
The transition we study is that between the dilute phase and the dense protein
phase which, in the present simplified picture, is completely analogous to the
vapor-liquid transition for particles interacting under the given pair
potential. Throughout this Section, the supersaturation will be defined as the
ratio of the density of the vapor phase to that of the vapor at coexistence,
$S\equiv\rho_{v}/\rho_{vc}$. The gradient coefficient, $K$, is calculated from
the pair potential using the approximation given in Ref.\cite{Lutsko2011a}
\begin{equation}
\beta K\simeq-\frac{2\pi}{45}d^{5}\beta v\left(  d\right)  +\frac{2\pi}%
{15}\int_{d}^{\infty}\left(  2d^{2}-5r^{2}\right)  \beta v\left(  r\right)
r^{2}dr
\end{equation}
where $d$ is the effective hard-sphere diameter for which we use the
Barker-Henderson approximation. For the temperature used here we get $\beta
K=1.80322\sigma^{5}$.

In the following, we will characterize the size of a cluster by either the
total excess number of molecules in the cluster,
\begin{equation}
\Delta N\equiv\int\left(  \rho(\mathbf{r})-\rho_{\infty}\right)  d\mathbf{r}%
\end{equation}
or by its equimolar radius, $R_{E}$, which is related to $\Delta N$ by
\begin{equation}
\Delta N=\frac{4\pi}{3}R_{E}^{3}(\rho_{0}-\rho_{\infty})
\end{equation}
For the simple capillary model one has that $\Delta N=\frac{4\pi}{3}R^{3}%
(\rho_{0}-\rho_{\infty})$ and $R_{E}=R$. For the extended model, a simple
calculation gives
\begin{equation}
\Delta N=\frac{\pi}{3w}\left(  R^{4}-(\max(R-w,0))^{4}\right)  (\rho_{0}%
-\rho_{\infty})
\end{equation}

\bigskip

\subsection{\bigskip Energy of formation of a cluster}

\begin{figure}
[ptb]\includegraphics[angle=-90,scale=0.25]{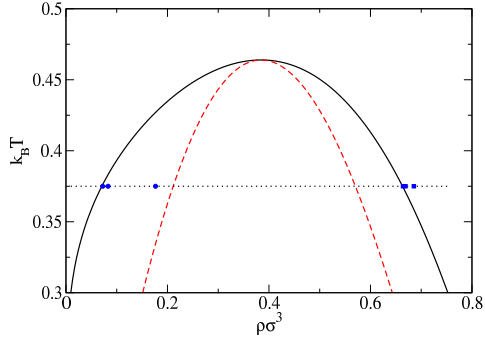}
\caption{The phase diagram for the
model globular protein. The full lines show the dilute/dense solution
coexistence, the broken lines are the spinodal and the horizontal line shows
the temperature used in the calculations.} \label{fig1}
\end{figure}

The dilute-solution/dense-solution phase diagram is shown in Fig. \ref{fig1}.
By definition, the coexistence concentrations correspond to saturation $S=1$
and at the spinodal the supersaturation is found to be $S=2.99$. We will
therefore illustrate the results of the models for supersaturations from
$S=1.025$ to $S=2.5$ corresponding, as will be seen, to quite large and very
small critical clusters, respectively.

\begin{figure}
[ptb]\includegraphics[angle=-90,scale=0.25]{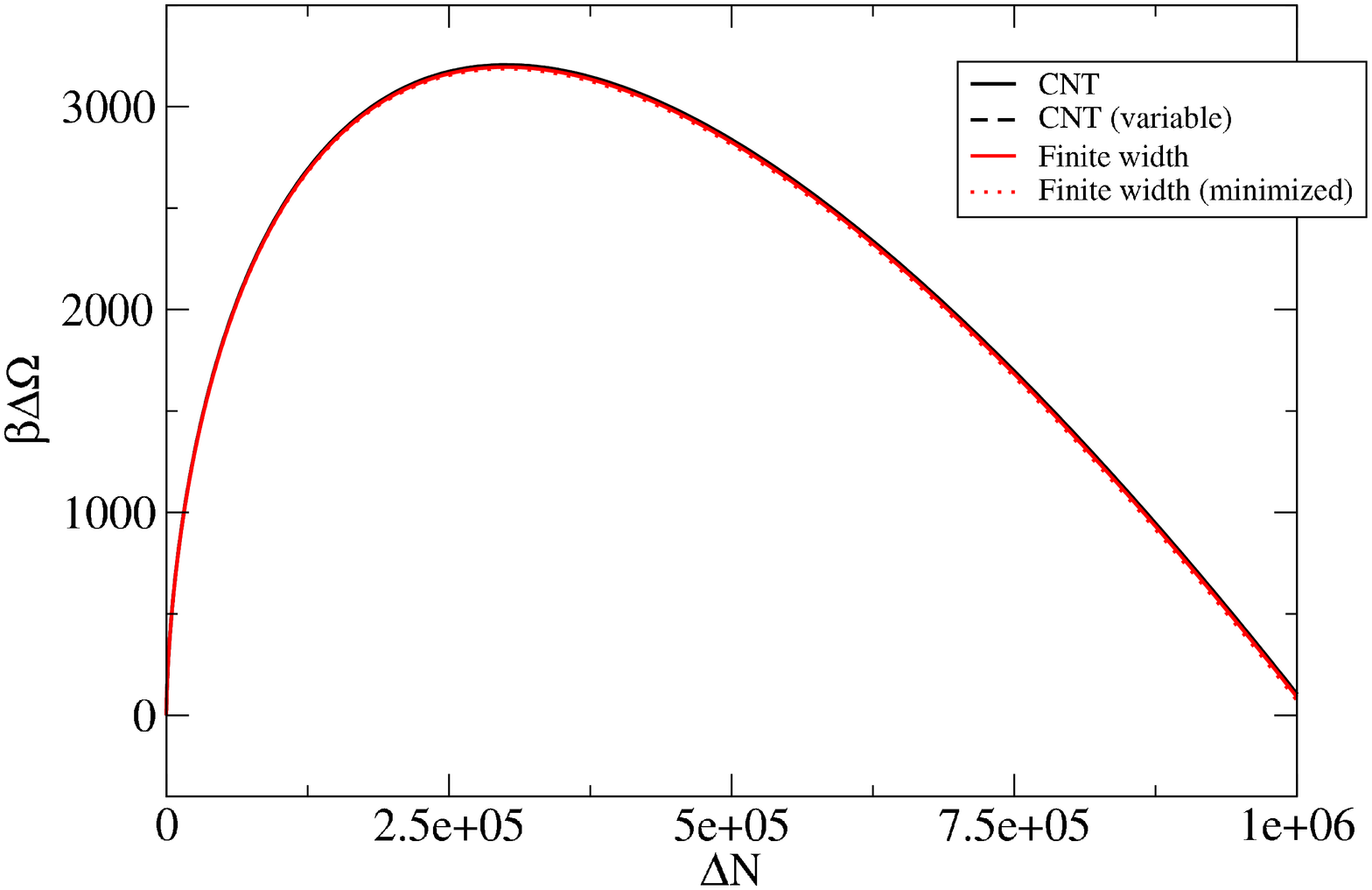}
\includegraphics[angle=-90,scale=0.25]{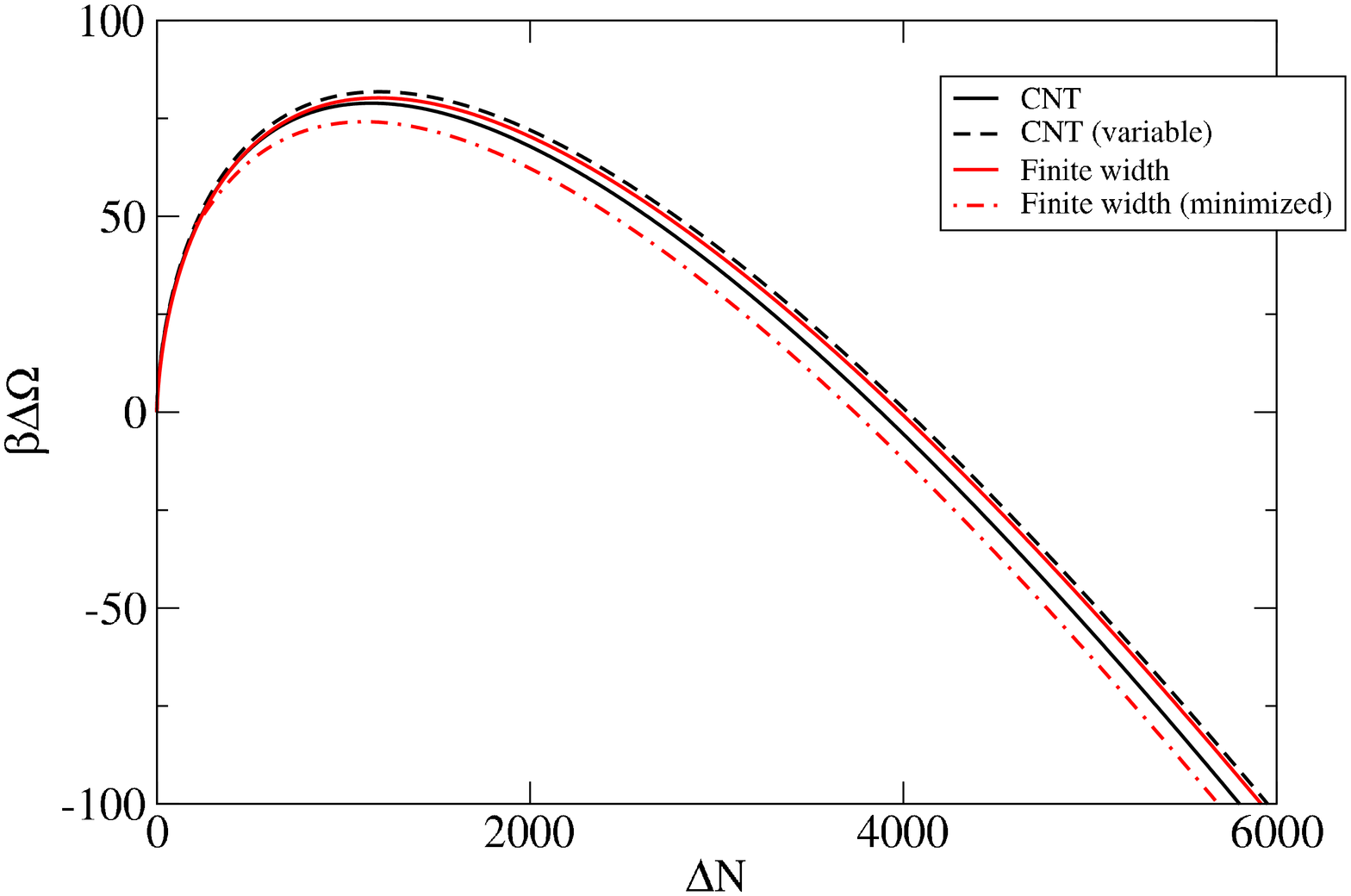}
\includegraphics[angle=-90,scale=0.25]{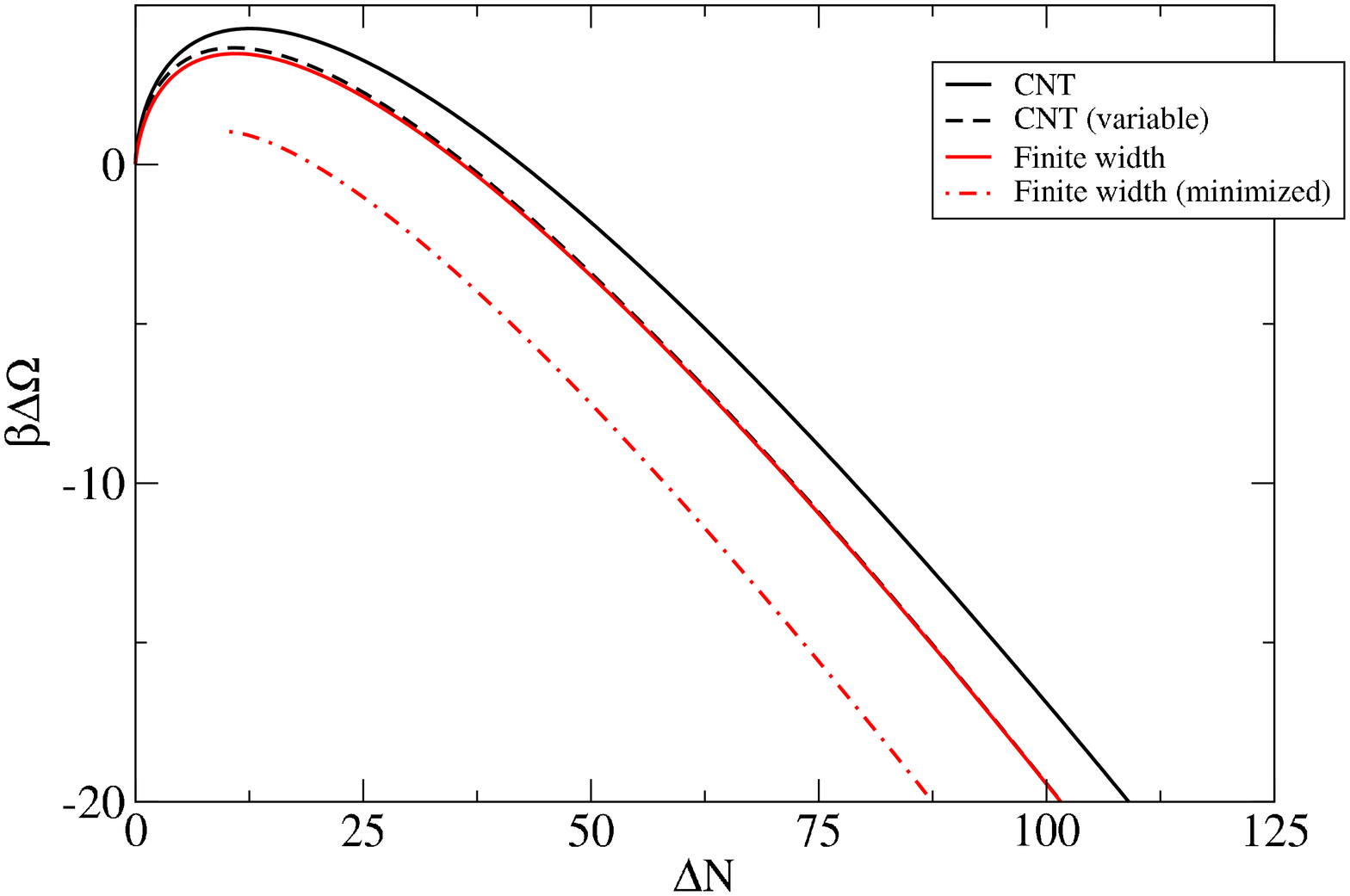} \caption{The free energy as a
function of cluster size (excess number of molecules in cluster) at $S=1.025, 1.175$ and $2.5$
as calculated using the capillary model with fixed $\gamma$, Eq.(\ref{cap1})
and Eq.(\ref{gan-fixed}), the capillary model with variable $\gamma$
(Eq.(\ref{b5})), the extended model with fixed width, Eq.(\ref{a1}) and
Eq.(\ref{width}), and the extended model minimized with respect to the width.}\label{fig2}%

\end{figure}

Figure \ref{fig2} shows the energy of formation of a cluster as a function of
its size for different values of supersaturation. The figures show the energy
as determined using CNT (capillary model with the value of $\gamma
=\gamma_{CNT}$ fixed to give the correct surface tension at coexistence
($S=1$), Eq.(\ref{gan-fixed})), the same model but with a
supersaturation-dependent value of $\gamma$ (Eq.(\ref{b5})), the extended
model with fixed width and the extended model minimized with respect to the
width. It is clear in all cases that the capillary model with fixed $\gamma$
gives lower free energies than when $\gamma$ is allowed to vary and that it is
also in closer agreement with the finite-width model. This is somewhat counter
intuitive. On the other hand, the extended model with fixed width gives
virtually the same results as when the energy is minimized with respect to the
width, showing that the simple fixed-width model is adequate. For these
reasons, only the capillary model with fixed $\gamma=\gamma_{CNT}$ and the
extended model with fixed width will be used below.



\subsection{The stationary distribution}

The exact stationary distribution is given in Eq.(\ref{stationary}). Although
both the capillary model and the extended model depend on a single variable, a
``radius'' denoted $R$ in both cases, the meaning of the parameter is not the
same. In order to make a meaningful comparison, we therefore give the
stationary distribution using the equimolar radius as the independent variable
and noting that
\begin{equation}
\bar{P}(R_{E}) = P(R) \frac{dR}{dR_{E}}.
\end{equation}
For the capillary model $R = R_{E}$ so that $\bar{P}(R_{E}) = P(R)$. For the
extended model, we find that
\begin{equation}
\frac{dR_{E}}{dR} = \frac{R^{3}-(max(R-w,0))^{3}}{3wR_{E}^{2}}.
\end{equation}

\begin{figure}
[ptb]\includegraphics[angle=-90,scale=0.25]{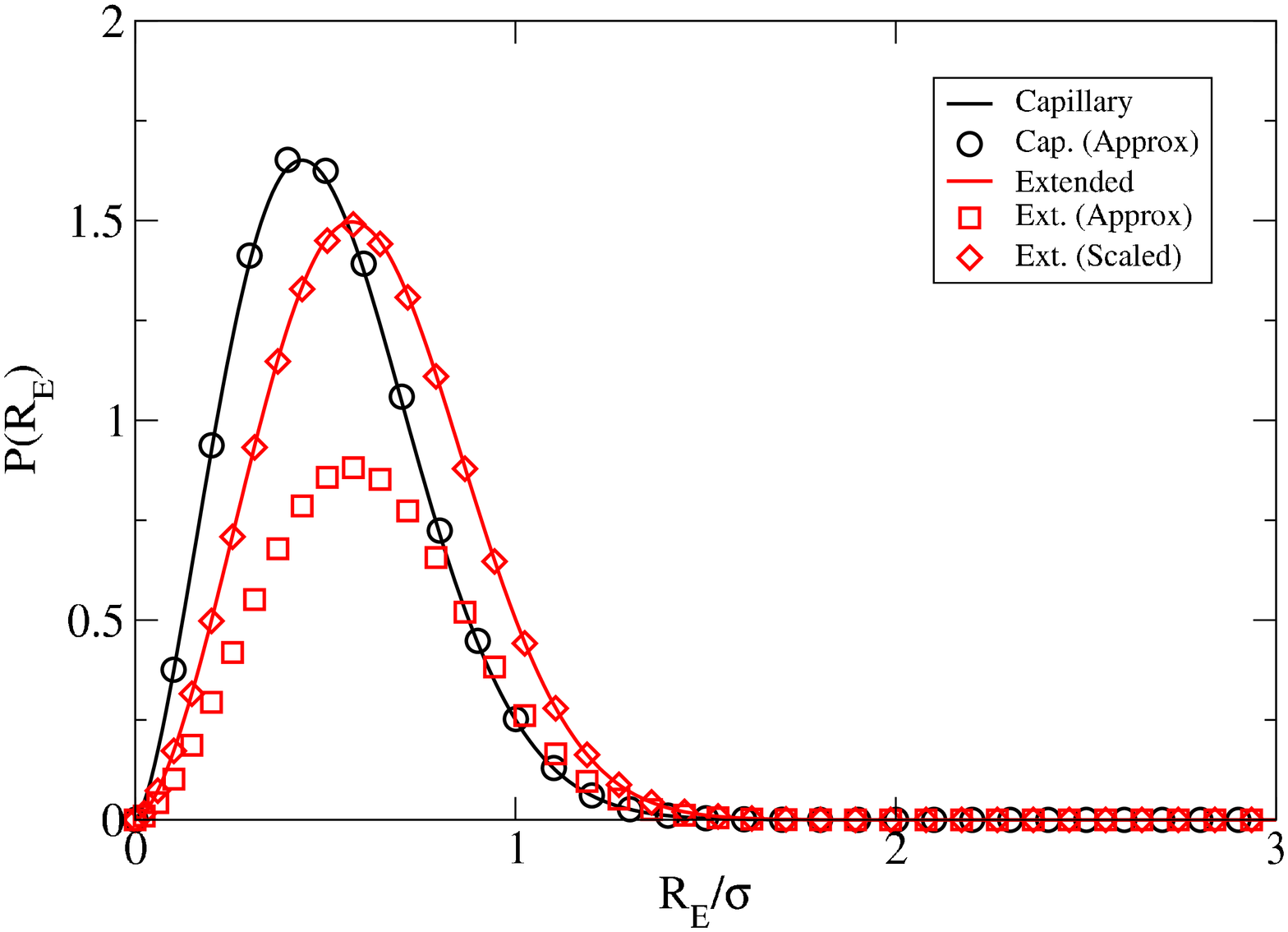}
\includegraphics[angle=-90,scale=0.25]{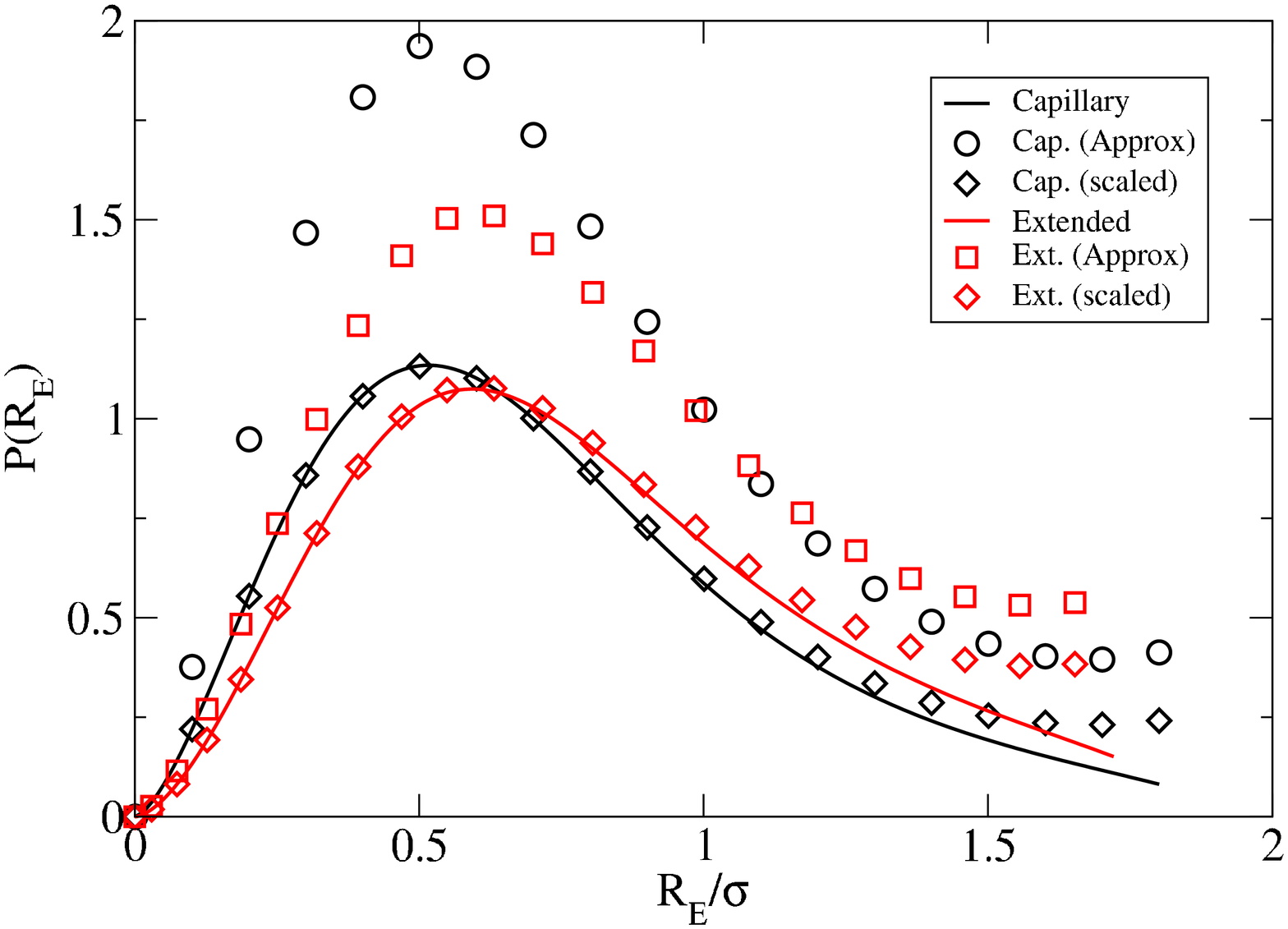}
\caption{The stationary
distribution for $S=1.025$, left panel, and $S=2.5$, right panel. Exact results, i.e. evaluations of Eq.(\ref{stationary}) are shown for the capillary and
extended models in units such that $\sigma=1$. Also shown is the approximation,
Eq.(\ref{PsApp}), and the approximate distribution scaled so as to have the same maximum as the exact result.}\label{fig5}%

\end{figure}

Figure \ref{fig5} shows the stationary distributions as calculated in CNT and
using the extended model for different values of the supersaturation. Also
shown are the approximate distribution used to evaluate the nucleation rate.
It is apparent that the approximate distribution has the right shape so that
it is indeed the case that the stationary distribution is well approximated by
the ``equilibrium'' distribution, $g(X)e^{-\beta\Delta\Omega}$. However, in
the case of the extended model, the approximate evaluation of the
normalization of the distribution is poor due to the rapidly changing analytic
structure of the free energy as a function of cluster radius for small
clusters. A surprising result is that the distributions for the capillary
model and for the extended model differ even for small supersaturation. This
is simply a reflection of the fact that the differences induced by the two
models are most pronounced for small clusters. regardless of the supersaturation.

\begin{figure}
[ptb]\includegraphics[angle=-90,scale=0.25]{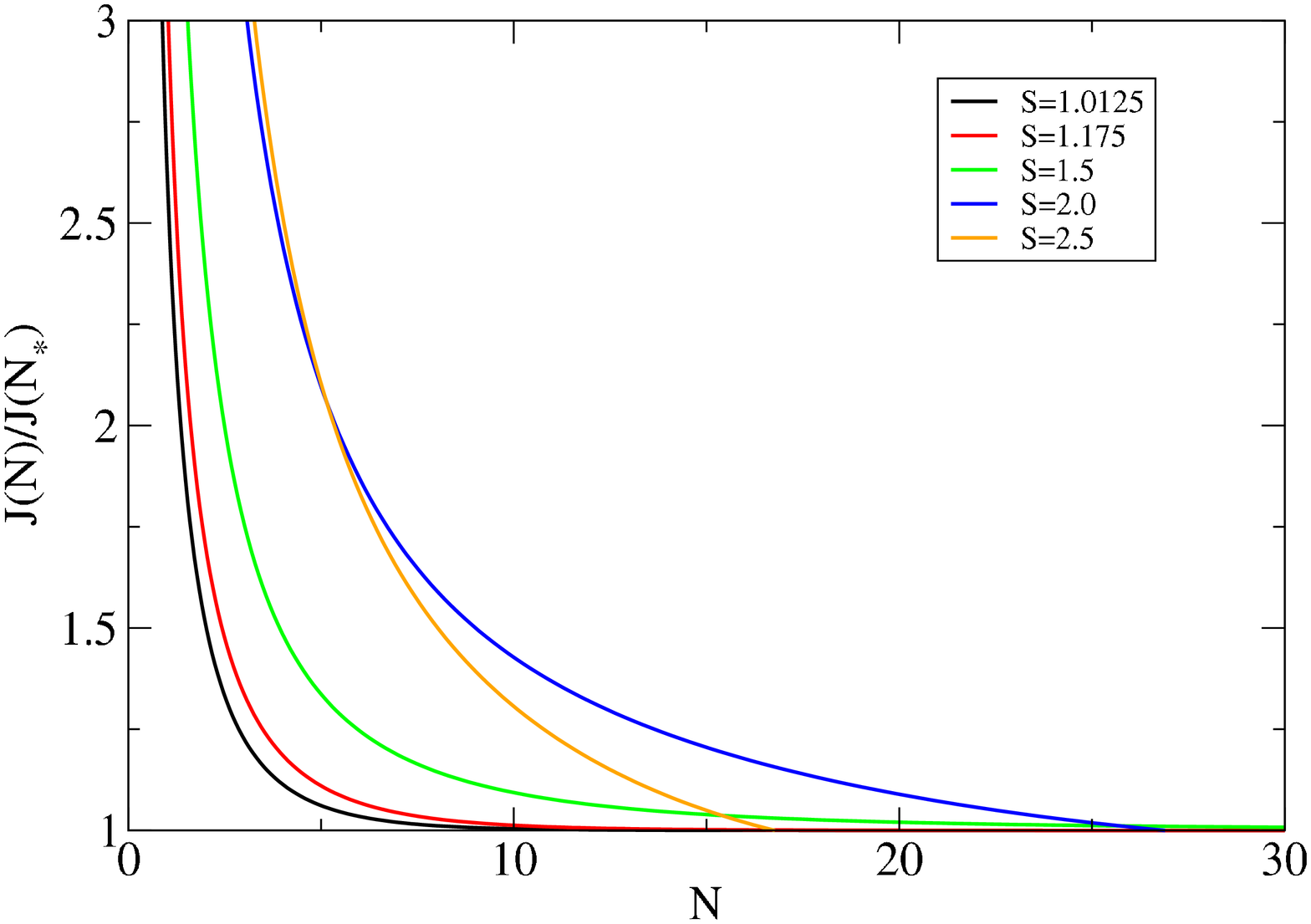}
\includegraphics[angle=-90,scale=0.25]{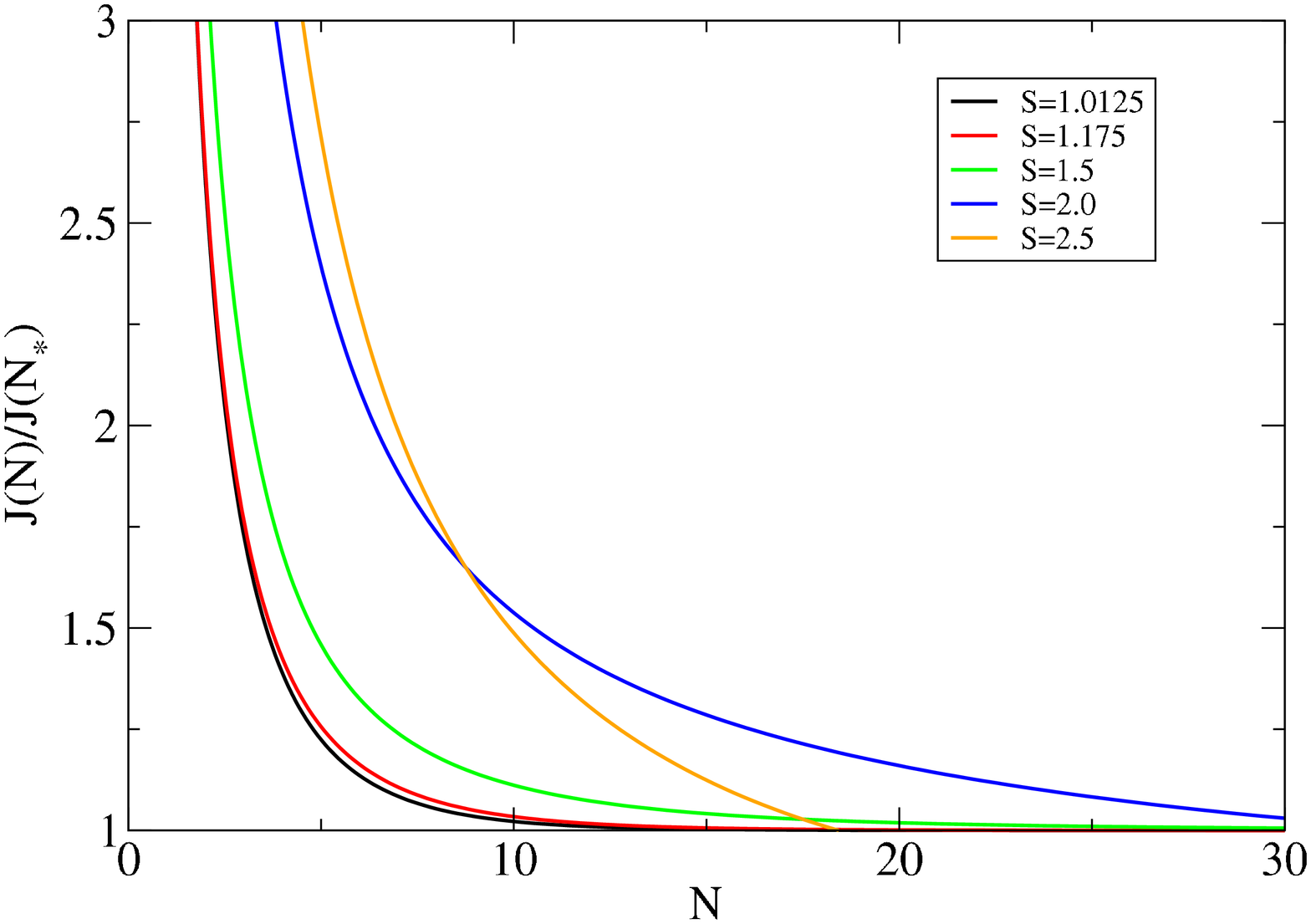}
\caption{The convergence of $J(N)$, Eq.(\ref{Jexact}), as a function of $N$ for the capillary model (left panel) and the extended model (right panel) for different values of the supersaturation. Note that in the case of the highest supersaturations, the nucleation rate does not reach a stable value before the integral extends up to the size of the critical cluster, $N_*$.}
\label{fig8}
\end{figure}

The approximate forms for the nucleation rates given in Eqs.(\ref{kcnt}) and
(\ref{kext}) are not calculated from the exact stationary distribution but,
rather, from the approximation given in Eq.(\ref{approx}). The figures show
that for CNT this approximation is quite good at low supersaturations but is
in considerable error for small clusters. The reason for the increasing error
is illustrated in Fig.\ref{fig8} which shows the convergence of the
\textquotedblleft exact\textquotedblright\ expression, Eq.(\ref{Jexact}), as a
function of the domain of integration. For large supersaturations, the
convergence is slow indicating that larger clusters are contributing
significantly to the evaluation of the nucleation rate. For large
supersaturations, this calls into question both the monomer attachment picture
that underlies CNT and the assumption of non-interacting clusters that is
tacitly used in both CNT and the dynamical theory. The main conclusion to be
drawn from this result is that even if one could accurately calculate the free
energy and distribution of clusters at large supersaturations, it is most
likely the case that cluster-cluster interactions would invalidate the
assumptions necessary to calculate the nucleation rate, both in CNT and in the
general framework described here.

\subsection{The nucleation rate}

Table \ref{tab1} gives the nucleation rate as calculated for the two models.
For the lowest supersaturation, the extended model gives a much higher
nucleation rate. This is almost entirely attributable to the fact that the
free energy of the critical cluster is about $10k_{B}T$ lower in the extended
model than in the capillary model (and, indeed, the log of the ratio of the
nucleation rates is about $9.7$). In all other cases, the nucleation rates of
the two models are comparable and similar to that given by the classical CNT formula.

\begin{table}
[tbp]%
\caption{Properties of the capillary and extended cluster models as functions of the Supersaturation, $S$. The classical nucleation rate, $J_{CNT}$ is given followed by
the critical equimolar radius, $R_{*E}$, the excess mass, $\Delta N_{*}$, and excess free energy, $\Delta \Omega_{*}$, of the critical cluster. Also given are the nucleation rates as calculated
from the ``exact'' expression, Eq.(\ref{Jexact}),  and the  approximate expression, Eq.(\ref{Japp}), $J$ and $J_{app}$, respectively. Lengths are in units of $\sigma$ and times are in units of $\sigma^2/D$. }\label{tab1}
\begin{ruledtabular}%
\begin{tabular}
[c]{c|c|c|c|c|c|c||c|c|c|c|c}%
&  & \multicolumn{5}{c}{Capillary} & \multicolumn{5}{c}{Extended Model} \\\hline
$S$  & $J_{CNT}e^{-\beta \Delta \Omega}$ & $R_{*E}$ & $\Delta N_{*}$ & $\Delta\beta\Omega_{*}$ & $\frac{J}{J_{CNT}}$ &
$\frac{J_{app}}{J_{CNT}}$ & $R_{*E}$ & $\Delta N_{*}$ & $\Delta\beta\Omega_{*}$ & $\frac{J}{J_{CNT}}$
& $\frac{J_{app}}{J_{CNT}}$ \\\hline
1.025  & $1.7 \times 10^{-4}$ & 49.5 & 300741 & 3205 & 0.4  & 0.37 &  49.5 & 300297 & 3195 & 6396 & 5564 \\
1.175  & $1.5 \times 10^{-3}$ & 7.8  & 1149   & 79   & 0.84 & 0.93 & 7.9   & 1188   &  80  & 0.09 & 0.11 \\
1.5    & $5.7 \times 10^{-3}$ & 3.27 &  83.6  & 14.0 & 1.03 & 1.45 &  3.39 & 92.9   & 14.9 & 0.22 & 0.40 \\
2.0    & $1.5 \times 10^{-2}$ & 2.1  & 21.6   & 5.9  & 0.7  & 1.9  & 2.2   & 22.9   &  5.9 & 0.5  & 1.6  \\
2.5    & $2.8 \times 10^{-2}$ & 1.8  & 12.6   & 4.3  & 0.9  & 2.2  & 1.7   & 11     &  3.5 & 1.1  & 3.8  \\
\end{tabular}
\end{ruledtabular}

\end{table}

\section{Conclusions}

In this paper an alternative approach to the semi-phenomenological description
of nucleation has been presented. The theory is represents a simplification of
a description of nucleation based on fluctuating hydrodynamics. One of the
advantages of this approach is that there is no ambiguity regarding the choice
of reaction coordinate. It also allows for the incorporation of more complex
cluster models. When evaluated using the capillary model for a cluster, the
results of Classical Nucleation Theory were recovered. A more realistic model
that includes a finite interfacial width was also developed. This gives rise
to sub-dominant corrections to the free energy of the type described by
Prestipino et al\cite{Prestipino}. It was also shown that the different
choices of reaction coordinate give rise to corrections to the free energy
that scale logarithmically with the cluster size. When the different models
were used to evaluate the nucleation rate, it was found that the dominant
effect was the difference in the free energy of the critical cluster in the
two models. Thus, despite the fact that the model for the structure of a
cluster affects all aspects of the dynamics in our general approach, it is
nevertheless the case, as is commonly assumed, that only the correction to the
free energy barrier is really important in determining the nucleation rate.
This provides post-hoc justification for the combination of the framework of
CNT with the use, e.g., of DFT to get better estimates for the free energy barrier.

We also were able to explicitly calculate the stationary distribution. It was
shown that the stationary distribution is well approximated, for small
clusters, by the ``equilibrium'' distribution. However, at high
supersaturations, it was found that the stationary distribution becomes quite
broad and that the nucleation rate becomes dependent on the domain of
integration over the cluster sizes. This indicates that the assumption of
cluster growth by monomer attachment and detachment is probably invalid as is
the assumption of non-interacting clusters. It is therefore the case that
quantitative prediction of nucleation rates is not really possible with the
theory discussed here. For the conditions considered here, the breakdown
occurs when the critical cluster contains between 20 and 100 molecules. A
cluster of 100 molecules is still very small and our results illustrate, in
some detail, the fact that Classical Nucleation Theory is remarkably robust
given the crude assumptions that go into it.

\begin{acknowledgments}
The work of JFL as partially supported in part by the European Space Agency
under contract number ESA AO-2004-070 and by FNRS Belgium under contract C-Net
NR/FVH 972. MD acknowledges support from the Spanish Ministry of Science and
Innovation (MICINN), FPI grant BES-2010-038422 (project AYA2009-10655).
\end{acknowledgments}

\appendix

\section{Approximate Nucleation Rate}

\label{AppRate} The nucleation rate is%
\begin{equation}
J^{-1}=\rho_{\infty}^{-1}D^{-1}\int_{0}^{X_{\ast}}N\left(  X\right)
g^{1/2}\left(  X\right)  \exp\left(  -\beta\Omega\left(  X\right)  \right)
\left(  \int_{X}^{X_{+}}g^{1/2}\left(  X^{\prime}\right)  \exp\left(
\beta\Omega\left(  X^{\prime}\right)  \right)  dX^{\prime}\right)  dX
\label{exact}%
\end{equation}
and is the exact expression for the nucleation rate given the assumptions of
the theory. In order to develop an approximation it is convenient to switch to
the canonical variable so that the nucleation rate becomes%
\begin{equation}
J^{-1}=\rho_{\infty}^{-1}D^{-1}\int_{0}^{Y_{\ast}}N\left(  Y\right)  \left(
-\beta\widetilde{\Omega}\left(  Y\right)  \right)  \left(  \int_{Y}^{Y_{+}%
}\exp\left(  \beta\widetilde{\Omega}\left(  Y^{\prime}\right)  \right)
dY^{\prime}\right)  dY
\end{equation}
The first step is to assume that the free energy has a maximum at $Y_{\ast}$
which is defined by%
\begin{equation}
\left.  \frac{\partial\widetilde{\Omega}\left(  Y\right)  }{\partial
Y}\right\vert _{Y_{\ast}}=0
\end{equation}
so that we can evaluate the inner integral by expanding%
\begin{equation}
\exp\left(  \beta\widetilde{\Omega}\left(  Y^{\prime}\right)  \right)
\sim\exp\left(  \beta\widetilde{\Omega}\left(  Y_{\ast}\right)  +\frac{1}%
{2}\beta\widetilde{\Omega}^{\prime\prime}\left(  Y_{\ast}\right)  \left(
Y^{\prime}-Y_{\ast}\right)  ^{2}\right)
\end{equation}
giving%
\begin{equation}
J^{-1}\sim\rho_{\infty}^{-1}D^{-1}\exp\left(  \beta\widetilde{\Omega}\left(
Y_{\ast}\right)  \right)  \sqrt{\frac{2\pi}{\left\vert \beta\widetilde{\Omega
}^{\prime\prime}\left(  Y_{\ast}\right)  \right\vert }}\int_{0}^{Y_{\ast}%
}N\left(  Y\right)  \exp\left(  -\beta\widetilde{\Omega}\left(  Y\right)
\right)  dY
\end{equation}
Before proceeding, note that this is really an approximation to the stationary
distribution of%
\begin{equation}
\widetilde{P}\left(  Y\right)  \sim JD^{-1}\exp\left(  \beta\widetilde{\Omega
}\left(  Y_{\ast}\right)  \right)  \sqrt{\frac{2\pi}{\left\vert \beta
\widetilde{\Omega}^{\prime\prime}\left(  Y_{\ast}\right)  \right\vert }}%
\exp\left(  -\beta\widetilde{\Omega}\left(  Y\right)  \right)  \label{approx}%
\end{equation}
or, in general,%
\begin{equation}
P\left(  X\right)  \sim JD^{-1}\exp\left(  \beta\Omega\left(  X_{\ast}\right)
\right)  \sqrt{\frac{2\pi g\left(  X\right)  g\left(  X_{\ast}\right)
}{\left\vert \beta\Omega^{\prime\prime}\left(  X_{\ast}\right)  \right\vert }%
}\exp\left(  -\beta\widetilde{\Omega}\left(  X\right)  \right)
\end{equation}
To further simplify, we assume that the free energy has a minimum at $Y=0$ and
that $\beta\widetilde{\Omega}\left(  Y\right)  =aY^{\alpha}+...$ and $N\left(
Y\right)  =N_{0}Y^{\beta}+...$ for some values of $\alpha,\beta>0$ giving%
\begin{align}
J^{-1}  &  \sim\rho_{\infty}^{-1}D^{-1}\exp\left(  \beta\widetilde{\Omega
}\left(  Y_{\ast}\right)  \right)  \sqrt{\frac{2\pi}{\left\vert \beta
\widetilde{\Omega}^{\prime\prime}\left(  Y_{\ast}\right)  \right\vert }}%
\int_{0}^{Y_{\ast}}N_{0}Y^{\beta}\exp\left(  -aY^{\alpha}\right)  dY\\
&  \sim\rho_{\infty}^{-1}D^{-1}\exp\left(  \beta\widetilde{\Omega}\left(
Y_{\ast}\right)  \right)  \sqrt{\frac{2\pi}{\left\vert \beta\widetilde{\Omega
}^{\prime\prime}\left(  Y_{\ast}\right)  \right\vert }}N_{0}\frac{1}{\alpha
}a^{-\frac{\beta+1}{\alpha}}\int_{0}^{\infty}Z^{\frac{\beta+1-\alpha}{\alpha}%
}\exp\left(  -Z\right)  dZ\nonumber\\
&  =\rho_{\infty}^{-1}D^{-1}\exp\left(  \beta\widetilde{\Omega}\left(
Y_{\ast}\right)  \right)  \sqrt{\frac{2\pi}{\left\vert \beta\widetilde{\Omega
}^{\prime\prime}\left(  Y_{\ast}\right)  \right\vert }}N_{0}\frac{1}{\alpha
}a^{-\frac{\beta+1}{\alpha}}\Gamma\left(  \frac{\beta+1}{\alpha}\right)
\nonumber
\end{align}
or%
\begin{equation}
J\sim D\rho_{\infty}\sqrt{\frac{1}{2\pi}}a^{\frac{1+\beta}{\alpha}}%
\allowbreak\frac{\alpha}{\Gamma\left(  \frac{1+\beta}{\alpha}\right)
}\left\vert \beta\widetilde{\Omega}^{\prime\prime}\left(  Y_{\ast}\right)
\right\vert ^{1/2}\exp\left(  -\beta\widetilde{\Omega}\left(  Y_{\ast}\right)
\right)  \label{rate}%
\end{equation}
It is easy to translate into the original variables to find%
\begin{equation}
J\sim D\rho_{\infty}\sqrt{\frac{1}{2\pi}}\allowbreak a^{\frac{1+\beta}{\alpha
}}\allowbreak\frac{\alpha}{\Gamma\left(  \frac{1+\beta}{\alpha}\right)
}\left\vert \beta\Omega^{\prime\prime}\left(  R_{\ast}\right)  g^{-1}\left(
R_{\ast}\right)  \right\vert ^{1/2}\exp\left(  -\beta\Omega\left(  R_{\ast
}\right)  \right)  . \label{final}%
\end{equation}

\section{Expansion of the free energy}

\label{Expansion}

Beginning with the model expression for the free energy,
\begin{align}
\Omega\left(  R;w\right)  -\Omega_{\infty} &  =\frac{4\pi}{3}\left(
R-w\right)  ^{3}\Delta\omega\left(  \rho_{0}\right)  \\
&  +2\pi\left(  2\overline{\omega}_{0}w+K\frac{\left(  \rho_{0}-\rho_{\infty
}\right)  ^{2}}{w}\right)  R^{2}\nonumber\\
&  -2\pi\left(  4\overline{\omega}_{1}w+K\frac{\left(  \rho_{0}-\rho_{\infty
}\right)  ^{2}}{w}\right)  Rw\nonumber\\
&  +2\pi\left(  2\overline{\omega}_{2}w+\frac{1}{3}K\frac{\left(  \rho
_{0}-\rho_{\infty}\right)  ^{2}}{w}\right)  w^{2}\nonumber
\end{align}
we minimize by setting the derivative with respect to the width, $w$, equal to
zero giving%
\begin{equation}
0=-2\left(  R-w\right)  ^{2}\Delta\omega\left(  \rho_{0}\right)  +\left(
2\overline{\omega}_{0}-K\frac{\left(  \rho_{0}-\rho_{\infty}\right)  ^{2}%
}{w^{2}}\right)  R^{2}-8\overline{\omega}_{1}wR+6\overline{\omega}_{2}%
w^{2}+\frac{1}{3}K\left(  \rho_{0}-\rho_{\infty}\right)  ^{2}%
\end{equation}
or, after dividing by $R^{3}$ and rearranging,%
\begin{align}
0 &  =\left(  \left(  -2\Delta\omega\left(  \rho_{0}\right)  +2\overline
{\omega}_{0}\right)  w^{2}-K\left(  \rho_{0}-\rho_{\infty}\right)
^{2}\right)  \\
&  +\left(  4\Delta\omega\left(  \rho_{0}\right)  -8\overline{\omega}%
_{1}\right)  w^{2}\frac{w}{R}\nonumber\\
&  +\left(  -2\Delta\omega\left(  \rho_{0}\right)  w^{2}+6\overline{\omega
}_{2}w^{2}+\frac{1}{3}K\left(  \rho_{0}-\rho_{\infty}\right)  ^{2}\right)
\left(  \frac{w}{R}\right)  ^{2}.\nonumber
\end{align}
We now expand in inverse powers of the radius,%
\begin{equation}
w=w_{0}\left(  1+a_{1}\frac{w_{0}}{R}+a_{2}\left(  \frac{w_{0}}{R}\right)
^{2}+...\right)
\end{equation}
and solve order by order in $\frac{w_{0}}{R}$ giving%
\begin{align}
w_{0}^{2} &  =\frac{K\left(  \rho_{0}-\rho_{\infty}\right)  ^{2}}{2\left(
-\Delta\omega\left(  \rho_{0}\right)  +\overline{\omega}_{0}\right)  }\\
a_{1} &  =\frac{\Delta\omega\left(  \rho_{0}\right)  -2\overline{\omega}_{1}%
}{\Delta\omega\left(  \rho_{0}\right)  -\overline{\omega}_{0}}\nonumber\\
a_{2} &  =\frac{1}{6\left(  \Delta\omega\left(  \rho_{0}\right)
-\overline{\omega}_{0}\right)  ^{2}}\left(  11\left(  \Delta\omega\left(
\rho_{0}\right)  \right)  ^{2}+\Delta\omega\left(  \rho_{0}\right)  \left(
5\overline{\omega}_{0}-60\overline{\omega}_{1}+9\overline{\omega}_{2}\right)
-9\overline{\omega}_{0\overline{\omega}2}-\overline{\omega}_{0}^{2}%
+60\overline{\omega}_{1}^{2}\right)  \nonumber
\end{align}
and%
\begin{align}
\Omega\left(  R;w\right)  -\Omega_{\infty}  & =\left(  \frac{4}{3}\pi
R^{3}\right)  \Delta\omega\left(  \rho_{0}\right)  +\left(  4\pi R^{2}\right)
\frac{K\left(  \rho_{0}-\rho_{\infty}\right)  ^{2}}{w_{0}}\\
& +\left(  4\pi R\right)  w_{0}^{2}\left(  2\Delta\omega\left(  \rho
_{0}\right)  -\overline{\omega}_{0}-2\overline{\omega}_{1}\right)  +O\left(
R^{0}\right)  \nonumber
\end{align}
Of course, we could solve the equation for the width exactly since it is
simply a fourth order polynomial in $w$,%
\begin{align}
0  & =-K\left(  \rho_{0}-\rho_{\infty}\right)  ^{2}+\left[  \frac{K\left(
\rho_{0}-\rho_{\infty}\right)  ^{2}}{3R^{2}}+2\overline{\omega}_{0}%
-2\Delta\omega\left(  \rho_{0}\right)  \right]  w^{2}\\
& +\left[  4\frac{\Delta\omega\left(  \rho_{0}\right)  -2\overline{\omega}%
_{1}}{R}\right]  w^{3}+\left[  \frac{6\overline{\omega}_{2}-2\Delta
\omega\left(  \rho_{0}\right)  }{R^{2}}\right]  w^{4}.\nonumber
\end{align}

We can also express the free energy in terms of the equimolar radius%
\begin{align}
R_{E}  & =\left(  \frac{1}{4w}\left(  R^{4}-(R-w)^{4}\right)  \right)
^{1/3}=R-\frac{1}{2}w_{0}+\frac{w_{0}}{12}\left(  1-6a_{1}\right)  \frac
{w_{0}}{R}\\
& +\frac{w_{0}}{24}\left(  4a_{1}-12a_{2}+1\right)  \left(  \frac{w_{0}}%
{R}\right)  ^{2}+O\left(  \frac{w_{0}}{R}\right)  ^{3}\nonumber
\end{align}
we find that%
\begin{equation}
R=R_{E}+\frac{w_{0}}{2}+\frac{w_{0}}{12}\left(  6a_{1}-1\right)  \frac{w_{0}%
}{R_{E}}+\frac{w_{0}\left(  6a_{2}-5a_{1}\right)  }{12}\left(  \frac{w_{0}%
}{R_{E}}\right)  ^{2}+O\left(  \left(  \frac{w_{0}}{R_{E}}\right)
^{3}\right)
\end{equation}
giving%
\begin{align}
\Omega\left(  R_{E};w\right)  -\Omega_{\infty} &  =\Delta\omega\left(
\rho_{0}\right)  \allowbreak\left(  \frac{4\pi}{3}R_{E}^{3}\right)
+w_{0}\left(  \frac{K\left(  \rho_{0}-\rho_{\infty}\right)  ^{2}}{w_{0}^{2}%
}+\frac{\Delta\omega\left(  \rho_{0}\right)  }{2}\right)  \left(  4\pi
R_{E}^{2}\right)  \\
&  +w_{0}^{2}\left(  6\frac{K\left(  \rho_{0}-\rho_{\infty}\right)  ^{2}%
}{w_{0}^{2}}+\left(  13+3a_{1}\right)  \Delta\omega-6\overline{\omega}%
_{0}-12\overline{\omega}_{1}\right)  \frac{2\pi}{3}R_{E}\allowbreak
+\allowbreak O\left(  R^{0}\right)  \allowbreak\nonumber
\end{align}

\bibliography{manuscript}

\bigskip
\end{document}